\documentclass[12pt,a4paper]{article}

\usepackage{epsfig}
\usepackage{amsmath,amsfonts,amssymb}
\usepackage{t1enc}
\usepackage{cite}

\providecommand{\openone}{\leavevmode\hbox{\small1\kern-3.8pt\normalsize1}}

\newcommand{\RE}{\text{Re}\,}
\newcommand{\IM}{\text{Im}\,}

\newcommand{\smn}{\sigma^{\mu \nu}}
\newcommand{\gmn}{g^{\mu \nu}}
\newcommand{\gns}{g^{\nu \sigma}}
\newcommand{\gm}{\gamma^\mu}
\newcommand{\gn}{\gamma^\nu}
\newcommand{\pam}{\partial_\mu}
\newcommand{\paM}{\partial^\mu}
\newcommand{\pan}{\partial_\nu}
\newcommand{\paN}{\partial^\nu}

\newcommand{\paND}{\overleftrightarrow{\partial^\nu}}
\newcommand{\DMD}{\overleftrightarrow{D^\mu}}
\newcommand{\Wmn}{W_{\mu \nu}}
\newcommand{\Wm}{W_\mu}
\newcommand{\Zmn}{Z_{\mu \nu}}
\newcommand{\Zm}{Z_\mu}
\newcommand{\Gn}{G_\nu^a}
\newcommand{\Gs}{G_\sigma^a}
\newcommand{\Amn}{A_{\mu \nu}}
\newcommand{\Bmn}{B_{\mu \nu}}
\newcommand{\Gmna}{G_{\mu \nu}^a}
\newcommand{\la}{\lambda^a}
\newcommand{\lb}{\lambda^b}

\newcommand{\Dsl}{D\!\!\!\!\!\!\not\,\,\,}
\newcommand{\qsl}{q\!\!\!\!\!\not\,\,}
\newcommand{\psl}{p\!\!\!\!\!\not\,\,}
\newcommand{\pwsl}{p_W\!\!\!\!\!\!\!\!\!\!\not\,\,\,\,\,\,\,}
\newcommand{\ptsl}{p_t\!\!\!\!\!\!\!\not\,\,\,\,}
\newcommand{\pbsl}{p_b\!\!\!\!\!\!\!\not\,\,\,\,}

\newcommand{\vl}{V_L}
\newcommand{\vr}{V_R}
\newcommand{\gl}{g_L}
\newcommand{\gr}{g_R}
\newcommand{\xttl}{X_{tt}^L}
\newcommand{\xttr}{X_{tt}^R}
\newcommand{\dvz}{d_V^Z}
\newcommand{\daz}{d_A^Z}
\newcommand{\dva}{d_V^\gamma}
\newcommand{\daa}{d_A^\gamma}
\newcommand{\dvG}{d_V^g}
\newcommand{\daG}{d_A^g}
\newcommand{\xctl}{X_{ct}^L}
\newcommand{\xctr}{X_{ct}^R}
\newcommand{\kl}{\kappa_{ct}^L}
\newcommand{\kr}{\kappa_{ct}^R}
\newcommand{\lL}{\lambda_{ct}^L}
\newcommand{\lR}{\lambda_{ct}^R}
\newcommand{\gcl}{\zeta_{ct}^L}
\newcommand{\gcr}{\zeta_{ct}^R}

\newcommand{\Al}{\mathcal{A}_L}
\newcommand{\Ar}{\mathcal{A}_R}
\newcommand{\Bl}{\mathcal{B}_L}
\newcommand{\Br}{\mathcal{B}_R}

\newcommand{\Alr}{\mathcal{A}_{L,R}}

\newcommand{\Blr}{\mathcal{B}_{L,R}}

\newcommand{\Fi}{f_i}
\newcommand{\Fj}{f_j}

\begin{document}

\begin{center}
\begin{Large}
{\bf A minimal set of top anomalous couplings}
\end{Large}

\vspace{0.5cm}
J. A. Aguilar--Saavedra  \\[0.2cm] 
{\it Departamento de Física Teórica y del Cosmos and CAFPE, \\
Universidad de Granada, E-18071 Granada, Spain} \\[0.1cm]
\end{center}

\begin{abstract}
We simplify the general form of the fermion-fermion-gauge boson interactions generated by dimension-six gauge-invariant effective operators by using the equations of motion to remove redundant operators. It is found that the most general vertex for off-shell fermions $\Fi$, $\Fj$ and an off-shell boson $V=W,Z,\gamma,g$ only involves $\gm$ and $\smn q_\nu$ terms, with $q=p_i-p_j$. Examples are given for the $Wtb$, $Ztt$, $\gamma tt$ and $gtt$ interactions, whose general expression is greatly simplified with respect to previous results in the literature. The same arguments apply to top flavour-changing neutral interactions with the $Z$ boson, the photon or the gluon, which can also be parameterised in full generality with only $\gm$ and $\smn q_\nu$ couplings. Explicit expressions are given for these vertices in terms of dimension-six gauge-invariant operators. We also discuss how effective operator coefficients might be determined 
from eventual measurements of anomalous couplings.
\end{abstract}

\section{Introduction}

The precise measurement of the couplings among the known fermions and bosons is a standard tool for the search of new physics beyond the Standard Model (SM). In particular, at the Large Hadron Collider (LHC), top quarks will be produced in large numbers, allowing to probe the top couplings with a great precision. Such a high precision is most welcome because, being the top the heaviest quark, effects of new physics on its couplings are expected to be larger than for any other fermion, and deviations with respect to the SM predictions might be detectable.
An adequate parameterisation of the most general interactions of the top quark (or any other fermion) with the gauge bosons is compulsory in order to search for new physics and to interpret the results of experimental measurements.
In particular, it is important to avoid the appearance of redundant parameters which only lead to a complication of the analyses, both from the theoretical and experimental side, without making them more general.

The on-shell interaction  between two fermions $\Fi$, $\Fj$ and a gauge boson $V=W,Z,\gamma,g$ can be parameterised in full generality as
\begin{eqnarray}
\mathcal{L}^\text{OS}_{V\Fi\Fj} & = & \bar \Fj \, \gamma^{\mu} \left( \Al
P_L + \Ar P_R
\right) \Fi \; V_\mu \nonumber \\
& & + \bar \Fj \, i \sigma^{\mu \nu} q_\nu 
\left( \Bl P_L + \Br P_R \right) \Fi \; V_\mu + \mathrm{H.c.} \,,
\label{ec:Vff}
\end{eqnarray}
where $q=p_i-p_j$ is the outgoing boson momentum and $\Alr$, $\Blr$ are form factors, which in general may depend on $q^2$. (For the flavour-conserving photon and gluon vertices $\Al = \Ar$ and for the flavour-changing ones $\Alr = 0$ due to gauge symmetry.)
A term proportional to $q^\mu$ does not give any contribution to the amplitudes for on-shell $V$, because in this case the vector boson polarisation $\epsilon_\mu$ satisfies $q^\mu \epsilon_\mu = 0$.\footnote{A $q^\mu$ term can also be dropped if $V$ couples to external massless fermions, in which case its contribution to the amplitude vanishes by application of the Dirac equation. This is indeed the case in several processes of interest at LHC and Tevatron, like for example single top production in $t$ and $s$ channels.}
Additional terms with different Lorentz structures can be brought into this form by using the on-shell conditions, namely, the Dirac equation.
For off-shell fermions $\Fi$, $\Fj$ the situation might seem quite different
because the Dirac equation cannot be used to restrict the number and structure of the Lagrangian terms. However, as we will show here, if the new anomalous couplings arise from dimension-six gauge-invariant effective operators, then the Lagrangian in 
Eq.~(\ref{ec:Vff}) is still the most general one. We recall here that effects of new physics at a high scale $\Lambda$ can be described by an effective Lagrangian~\cite{Burges:1983zg,Leung:1984ni,Buchmuller:1985jz}
\begin{equation}
\mathcal{L}^\text{eff} = \sum \frac{C_x}{\Lambda^2} O_x + \dots \,,
\label{ec:effL}
\end{equation}
where $O_x$ are dimension-six gauge-invariant operators and $C_x$ are complex constants. (Higher-order corrections from operators of higher dimension, suppressed by higher powers of $\Lambda$, are neglected in this work.) Among the operators listed in Ref.~\cite{Buchmuller:1985jz}, fourteen contribute to top electroweak anomalous couplings,
\begin{align}
& O_{\phi q}^{(3,ij)} = i (\phi^\dagger \tau^I D_\mu  \phi)
       (\bar q_{Li} \gm \tau^I q_{Lj}) \,,
  && O_{Du}^{ij} = (\bar q_{Li} \,D_\mu u_{Rj}) \, 
       D^\mu \,\tilde \phi \,, \notag \\
& O_{\phi q}^{(1,ij)} = i (\phi^\dagger D_\mu \phi) (\bar q_{Li} \gm q_{Lj}) \,,
  && O_{\bar Du}^{ij} = (D_\mu  \bar q_{Li} \,u_{Rj}) \,
        D^\mu \,\tilde \phi \,, \notag \\
& O_{\phi \phi}^{ij} = i (\tilde \phi^\dagger D_\mu \phi)
        (\bar u_{Ri} \gm d_{Rj}) \,,
  && O_{Dd}^{ij} = (\bar q_{Li} \,D_\mu d_{Rj}) \, D^\mu \, \phi \,, \notag \\
& O_{\phi u}^{ij} = i (\phi^\dagger D_\mu \phi) (\bar u_{Ri} \gm u_{Rj}) \,,
  && O_{\bar Dd}^{ij} = (D_\mu  \bar q_{Li} \, d_{Rj}) \,
        D^\mu \, \phi \,, \notag \\
& O_{uW}^{ij} = (\bar q_{Li} \smn \tau^I u_{Rj}) \tilde \phi \, \Wmn^I \,,
  && O_{qW}^{ij} = \bar q_{Li} \gm \tau^I D^\nu q_{Lj} \Wmn^I \,, \notag \\
& O_{dW}^{ij} = (\bar q_{Li} \smn \tau^I d_{Rj}) \phi \, \Wmn^I \,,
  && O_{qB}^{ij} = \bar q_{Li} \gm D^\nu q_{Lj} \Bmn \,, \notag \\
& O_{uB\phi}^{ij} = (\bar q_{Li} \smn u_{Rj}) \tilde \phi \, \Bmn \,,
  && O_{uB}^{ij} = \bar u_{Ri} \gm D^\nu u_{Rj} \Bmn \,,
\label{ec:Oall}
\end{align}
up to different values of the flavour indices $i,j=1,2,3$. Here $\bar q_{Li}$, $u_{Ri}$ and $d_{Ri}$ are the quark fields in standard notation (for details see the next section). Operators with $i=j=3$ contribute to the $Wtb$, $Ztt$ or $\gamma tt$ vertices, while operators
involving two up-type quarks with $i,j=1,3/3,1$ or $i,j=2,3/3,2$ contribute to flavour-changing neutral (FCN) top-up and top-charm interactions, respectively. Only three operators
(up to flavour indices) contribute to strong interactions,
\begin{align}
& O_{uG\phi}^{ij} = (\bar q_{Li} \la \smn u_{Rj}) \tilde \phi \, \Gmna \,,
  && O_{qG}^{ij} = \bar q_{Li} \la  \gm D^\nu q_{Lj} \Gmna \,, \notag \\
& && O_{uG}^{ij} = \bar u_{Ri} \la \gm D^\nu u_{Rj} \Gmna \,.
\label{ec:Ostr}
\end{align}
For $i=j=3$ they give diagonal $gtt$ couplings whereas for $i \neq j$ the interactions are flavour-changing, as the electroweak ones.

All operators in the left columns of Eqs.~(\ref{ec:Oall}), (\ref{ec:Ostr}) yield $\gm$ and $\smn q_\nu$ terms, while those in the right columns give $k^\mu \equiv (p_i+p_j)^\mu$ and $q^\mu$ terms or more complicated Lorentz structures.
Not all these contributions to top couplings are independent, however. In Ref.~\cite{Grzadkowski:2003tf} it was pointed out that $O_{qW}^{33}$, $O_{qB}^{33}$, $O_{uB}^{33}$ are redundant and can be expressed in terms of other operators in Eqs.~(\ref{ec:Oall}) with $i=j=3$, plus four-fermion interactions. This implies in particular that their contributions to the $Wtb$, $Ztt$ and $\gamma tt$ couplings can be expressed in terms of other operator contributions  both for on-shell and off-shell external particles. 
Here we will generalise this result for operators with $i \neq j$, including also strong interactions. We will find expressions which allow to write: (i) $O_{qW}^{ij}$, $O_{qB}^{ij}$ and $O_{uB}^{ij}$ in terms of operators in the left column of Eqs.~(\ref{ec:Oall}) plus four-fermion interactions, extending the results in Ref.~\cite{Grzadkowski:2003tf} to the case of $i \neq j$; (ii) $O_{qG}^{ij}$ and $O_{uG}^{ij}$ in terms of $O_{uG\phi}^{ij}$ plus four-fermion interactions. After proving that these operators are redundant, they can be excluded from further consideration in the same way as several other gauge-invariant dimension-six redundant operators one may construct~\cite{Buchmuller:1985jz} are not considered.

Concerning the remaining operators, it has been previously noted that using the equations of motion $O_{\bar Du}^{33}$ and $O_{\bar Dd}^{33}$ can be expressed in terms of $O_{Du}^{33}$, $O_{Dd}^{33}$, respectively, plus additional terms.
More recently, in Ref.~\cite{AguilarSaavedra:2008gt} it has been shown with a direct calculation of the amplitudes that, for the specific case of the $Wtb$ vertex, the contributions of $O_{Du}^{33}$, $O_{\bar Du}^{33}$, $O_{Dd}^{33}$ and $O_{\bar Dd}^{33}$ can be rewritten in terms of $\gm$ and $\smn q_\nu$ terms. Here we will show that the underlying reason for this simplification is that
the four operators $O_{Du}^{ij}$, $O_{\bar Du}^{ij}$, $O_{Dd}^{ij}$ and $O_{\bar Dd}^{ij}$ are actually redundant. Therefore, only the operators in the left columns of Eqs.~(\ref{ec:Oall}), (\ref{ec:Ostr}) are independent, and all of them give $\gm$ or $\smn q_\nu$ contributions to the vertices.

We must emphasise here that the fundamental principle which allows to rewrite vertex contributions into $\gm$ and $\smn q_\nu$ terms for off-shell particles is gauge symmetry. It is well-known that for off-shell fermions the Lorentz structure in Eq.~(\ref{ec:Vff}) is not the most general one and, for example, a $k^\mu$ term cannot be rewritten into $\smn q_\nu$ plus $\gm$ terms using the Gordon identities if the fermions are off-shell. But when these trilinear terms are generated from gauge-invariant operators, they have associated quartic interactions (among several others) which also contribute to the amplitudes. In this way, the contribution of a $k^\mu$ term {\em plus} the additional contributions related by gauge symmetry are equivalent to the one from a combination of $\smn q_\nu$ and $\gm$ terms. This fact will be explicitly shown here with examples of amplitude calculations.

The aim of this paper is to find general expressions of the electroweak top anomalous interactions generated by dimension-six gauge-invariant operators, which are also minimal in the sense that they involve a set of couplings as small as possible.
Section~\ref{sec:2} is devoted to obtain relations which will eventually prove that the operators listed in the right columns of Eqs.~(\ref{ec:Oall}), (\ref{ec:Ostr}) are redundant and may be safely excluded.
The results obtained are then applied in section~\ref{sec:3} to the $Wtb$, $Ztt$, $\gamma tt$ and $gtt$ vertices. We find much simpler vertex structures in comparison to previous works~\cite{Gounaris:1996vn,Gounaris:1996yp,Whisnant:1997qu,Yang:1997iv}. 
Altogether, these interactions can be described with only twelve independent anomalous couplings which are the coefficients of the effective $\gm$ and $\smn q_\nu$ interactions. The eventual measurement of these anomalous couplings might be used to determine effective operator coefficients.
In section~\ref{sec:4} we present results for $Ztu/Ztc$, $\gamma tq/\gamma tc$ and $gtu/gtc$ interactions, simplifying previous results~\cite{Ferreira:2008cj,Coimbra:2008qp}. For example, the $Ztu$ and $Ztc$ interactions can be each described by only four quantities and the $\gamma tu$ and $\gamma tc$ vertices only need two parameters, which are the most convenient ones to express observables such as cross sections and branching ratios for FCN decays. As we will see, these parameters turn out to be independent, despite the gauge relation between the $Z$ boson and photon fields.
In appendix~\ref{sec:c} we give
explicit examples to show how gauge symmetry ensures that the contributions to the amplitudes are the same in all cases when operators are rewritten.
In appendix~\ref{sec:a} we collect the effective operator contributions to the $Wtb$, $Ztt$, $\gamma tt$ and $gtt$ vertices, while in appendix~\ref{sec:b} we do the same for the flavour-changing ones.

\section{Effective operator equalities}
\label{sec:2}

In this paper
we follow the notation of Ref.~\cite{Buchmuller:1985jz} for gauge invariant effective operators with slight normalisation changes and sign differences, also introducing flavour indices. We denote by
\begin{equation}
q_{Li} = \left(\! \begin{array}{c} u_{Li} \\ d_{Li} \end{array} \!\right) \;,\quad
 u_{Ri} \;,\quad d_{Ri} \quad \quad (i=1,2,3)
\end{equation}
the quark weak interaction eigenstates, with $(u_1,u_2,u_3)=(u,c,t)$ and
$(d_1,d_2,d_3)=(d,s,b)$ in the usual notation. Analogously, 
$\ell_{Li}$ and $e_{Ri}$ the lepton doublets and singlets, respectively.
The covariant derivative is
\begin{equation}
D_\mu = \partial_\mu + i g_s \frac{\la}{2} G_\mu^a
+ i g \frac{\tau^I}{2} W_\mu^I + i g' Y B_\mu \,,
\end{equation}
where $G_\mu^a$, $W_\mu^I$ and $B_\mu$ are the gauge fields for
$\mathrm{SU}(3)$, $\mathrm{SU}(2)_L$ and $\mathrm{U}(1)_Y$,
$\la$ are the Gell-Mann matrices with $a=1\dots 8$,
$\tau^I$ the Pauli matrices for $I=1,2,3$ and
$Y$ is the hypercharge (with $Q=T_3+Y$) of the field to which $D_\mu$ is applied.
The charged $W$ boson fields are
\begin{equation}
W_\mu^\pm = \frac{1}{\sqrt 2} \left( W_\mu^1 \mp i W_\mu^2 \right)
\end{equation}
and the $Z$ and photon are related to the $W^3$, $B$ fields by
\begin{align}
& Z_\mu = c_W W_\mu^3 - s_W B_\mu \,, \notag \\
& A_\mu = s_W W_\mu^3 + c_W B_\mu \,,
\end{align}
where $s_W$ and $c_W$ are the sine and cosine of the weak angle $\theta_W$, respectively. The field strength tensors for $\text{SU}(3)$, $\text{SU}(2)_L$ and $\text{U}(1)_Y$ are
\begin{align}
& G_{\mu \nu}^a = \partial_\mu G_\nu^a - \partial_\nu G_\mu^a - g_s f_{abc} G_\mu^a G_\nu^c \,, \notag \\
& W_{\mu \nu}^I = \partial_\mu W_\nu^I - \partial_\nu W_\mu^I - g \epsilon_{IJK} W_\mu^J W_\nu^K \,, \notag \\
& B_{\mu \nu} = \partial_\mu B_\nu - \partial_\nu B_\mu \,,
\end{align}
and the dual tensors are
\begin{equation}
\tilde F^{\mu \nu} = \frac{1}{2} \epsilon^{\mu \nu \tau \rho} F_{\tau \rho} \,,
\end{equation}
for $F=B,W^I,G^a$ with $\epsilon_{0123}=1$. The SM Higgs doublet $\phi$ has vacuum expectation value
\begin{equation}
\langle \phi \rangle = \frac{1}{\sqrt 2}
\left(\! \begin{array}{c} 0 \\ v \end{array} \!\right) \,,
\end{equation}
with $v=246$ GeV, and we define $\tilde \phi = \epsilon \phi^*$, $\epsilon = i \tau^2$. We will use 
the dimension-four equations of motion of the quark fields
\begin{align}
& i \Dsl q_{Li} = Y^u_{ij} \, u_{Rj} \tilde \phi + Y^d_{ij} \, d_{Rj} \phi \,, \notag \\
& i \Dsl u_{Ri} = Y^{u\dagger}_{ij} {\tilde \phi}^\dagger q_{Lj} \,, \notag \\
& i \Dsl d_{Ri} = Y^{d\dagger}_{ij} \phi^\dagger q_{Lj} \,,
\label{ec:EMq}
\end{align}
with $Y^u$, $Y^d$ the $3 \times 3$ matrices of up- and down-type quark Yukawa couplings.
The equations of motion of gauge fields are
\begin{eqnarray}
\pan B^{\nu \mu} & - & g' \left\{ -\frac{1}{2} \bar \ell_{Li} \gm \ell_{Li} - \bar e_{Ri} \gm e_{Ri} + \frac{1}{6} \bar q_{Li} \gm q_{Li} + \frac{2}{3} \bar u_{Ri} \gm u_{Ri} - \frac{1}{3} \bar d_{Ri} \gm d_{Ri} \right. \notag \\
& & \left. +\frac{i}{2} \phi^\dagger \DMD \phi \right\} = 0 \,, \notag \\
(D_\nu W^{\nu \mu})^I & - & g \left\{ \bar \ell_{Li} \gm \frac{\tau^I}{2} \ell_{Li} +
\bar q_{Li} \gm \frac{\tau^I}{2} q_{Li} + i \left[ \phi^\dagger \frac{\tau^I}{2} D^\mu \phi - (D^\mu \phi)^\dagger \frac{\tau^I}{2} \phi \right] \right\} = 0 \,, \notag \\
(D_\nu G^{\nu \mu})^a & - & g_s \left\{ \bar q_{Li} \frac{\la}{2} \gm q_{Li} +
\bar u_{Ri} \frac{\la}{2} \gm u_{Ri} + \bar d_{Ri} \frac{\la}{2} \gm d_{Ri} \right\} = 0 \,,
\end{eqnarray}
summing over flavours $i=1,2,3$, with
\begin{eqnarray}
(D_\mu W_{\nu \sigma})^I & = & \pam W_{\nu\sigma}^I - g \epsilon_{IJK} W_\mu^J W_{\nu\sigma}^K \,, \notag \\
(D_\mu G_{\nu \sigma})^a & = & \pam G_{\nu\sigma}^a - g_s f_{abc} G_\mu^b G_{\nu\sigma}^c \,.
\end{eqnarray}
The equation of motion for the scalar field is
\begin{equation}
D_\mu D^\mu \phi - m^2 \phi + \lambda (\phi^\dagger \phi) \phi + Y^{e\dagger}_{ij} \bar e_{Ri} \ell_{Lj} + Y^u_{ij} (\bar q_{Li} \epsilon)^T u_{Rj}
+ Y^{d\dagger}_{ij} \bar d_{Ri} q_{Lj} \,,
\end{equation}
and we will also use
\begin{equation}
D_\mu D^\mu \tilde \phi - m^2 \tilde \phi + \lambda (\tilde \phi^\dagger \tilde \phi) \tilde \phi + Y^e_{ij} (\bar \ell_{Li} \epsilon)^T \bar e_{Rj} + Y^{u\dagger}_{ij} \bar u_{Ri} q_{Lj}
+ Y^d_{ij} (\bar q_{Li} \epsilon)^T d_{Rj} \,.
\end{equation}
In the rest of this section we will use the dimension-four equations of motion for the interacting SM fields to obtain relations among the effective operators. These equations can be used to remove redundant operators even for off-shell external particles~\cite{Arzt:1993gz}. 
Although for definiteness we restrict ourselves to operators involving quarks, it is evident that the same arguments apply to the lepton sector where the fields have the same isospin structure but are singlets under $\mathrm{SU}(3)$, and the equivalent leptonic operators are redundant as well. In appendix~\ref{sec:c} we provide examples showing that the rewriting of contributions implied by the operator equalities give the same result in amplitude calculations even with off-shell fermions and bosons.

\subsection{Equalities for $O_{qW}^{ij}$, $O_{qB}^{ij}$, $O_{uB}^{ij}$, $O_{qG}^{ij}$ and $O_{uG}^{ij}$}

In Ref.~\cite{Grzadkowski:2003tf} it was pointed out that $O_{qW}^{33}$, $O_{qB}^{33}$ and $O_{uB}^{33}$ are redundant and can be written in terms of other gauge-invariant operators. We extend this result to the non-diagonal case $i \neq j$ also including $O_{qG}^{ij}$ and $O_{uG}^{ij}$. The desired relations can be obtained by writing all these operators as
\begin{equation}
O_x^{ij} = \frac{1}{2} \left[ O_x^{ij} + (O_x^{ji})^\dagger) \right]
 + \frac{1}{2} \left[ O_x^{ij} - (O_x^{ji})^\dagger \right] \,,
\end{equation}
and relating each of the terms between brackets to other gauge invariant operators. Note that for $i=j$ the first term is hermitian while the second one is anti-hermitian.
For the first one we have
\begin{eqnarray}
O_{qW}^{ij} + (O_{qW}^{ji})^\dagger & = & (\bar q_{Li} \gm \tau^I q_{Lj}) \, (D^\nu W_{\nu \mu})^I \,, \notag \\
O_{qB}^{ij} + (O_{qB}^{ji})^\dagger & = & (\bar q_{Li} \gm q_{Lj}) \, \paN B_{\nu \mu} \,, \notag \\
O_{uB}^{ij} + (O_{uB}^{ji})^\dagger & = & (\bar u_{Ri} \gm u_{Rj}) \, \paN B_{\nu \mu} \,, \notag \\
O_{qG}^{ij} + (O_{qG}^{ji})^\dagger & = & (\bar q_{Li} \la \gm q_{Lj}) \, (D^\nu G_{\nu \mu})^a \,, \notag \\
O_{uG}^{ij} + (O_{uG}^{ji})^\dagger & = & (\bar u_{Ri} \la \gm u_{Rj}) \, (D^\nu G_{\nu \mu})^a \,,
\end{eqnarray}
up to a total derivative.
These sums can then be transformed using the gauge field equations of motion. For the second term we make use of the operators involving dual field strengths, which we define with an extra $i$ factor,
\begin{align}
& O_{q\tilde W}^{ij} = i \bar q_{Li} \gm D^\nu \tau^I q_{Lj} \tilde W_{\mu \nu}^I \,,
&& O_{q\tilde G}^{ij} = i \bar q_{Li} \la \gm D^\nu q_{Lj} \tilde G_{\mu \nu}^a \,, \notag \\
& O_{q\tilde B}^{ij} = i \bar q_{Li} \gm D^\nu q_{Lj} \tilde B_{\mu \nu} \,,
&& O_{u\tilde G}^{ij} = i \bar u_{Ri} \la \gm D^\nu u_{Rj} \tilde G_{\mu \nu}^a \,,
 \notag \\
& O_{u\tilde B}^{ij} = i \bar u_{Ri} \gm D^\nu u_{Rj} \tilde B_{\mu \nu} \,.
\end{align}
Their relation with the ones involving $B$, $W^I$, $G^a$,
\begin{eqnarray}
O_{q \tilde W}^{ij} & = & O_{qW}^{ij} + \frac{1}{2} \bar q_{Li} \smn \tau^I i \Dsl q_{Lj} \Wmn^I \,, \notag \\
O_{q \tilde B}^{ij} & = & O_{qB}^{ij} + \frac{1}{2} \bar q_{Li} \smn i \Dsl q_{Lj} \Bmn \,, \notag \\
O_{u \tilde B}^{ij} & = & -O_{uB}^{ij} - \frac{1}{2} \bar u_{Ri} \smn i \Dsl u_{Rj} \Bmn \,, \notag \\
O_{q \tilde G}^{ij} & = & O_{qG}^{ij} + \frac{1}{2} \bar q_{Li} \la \smn i \Dsl q_{Lj} \Gmna \,, \notag \\
O_{u \tilde G}^{ij} & = & -O_{uG}^{ij} + \frac{1}{2} \bar u_{Ri} \la \smn i \Dsl u_{Rj} \Gmna \,,
\label{ec:OOdual}
\end{eqnarray}
can be trivially obtained from the equality~\cite{Buchmuller:1985jz}
\begin{equation}
\tilde F^{\mu \nu} \gamma_\mu D_\nu \psi_\pm = \pm \left(
i F^{\mu \nu} \gamma_\mu D_\nu - \frac{1}{2} F^{\mu \nu} \sigma_{\mu \nu} \Dsl \right) \psi_\pm \,,
\label{ec:Fbuch}
\end{equation}
where the spinors $\psi_\pm$ satisfy $\gamma_5 \psi_\pm = \pm \psi_\pm$.\footnote{In Refs.~\cite{Whisnant:1997qu,Yang:1997iv} relations equivalent to those in Eqs.~(\ref{ec:OOdual}) plus the hermitian conjugate are quoted but without the $1/2$ factors in the terms with the $\smn$ matrices. In order to clarify this discrepancy, we have confirmed Eqs.~(\ref{ec:OOdual}) with a direct calculation using the property $\smn \gamma^\sigma = - \epsilon^{\mu \nu \sigma \rho} \gamma_5 \gamma_\rho + i \gm g^{\nu \sigma} - i \gamma^\nu g^{\mu \sigma}$.
Notice also a different sign in Ref.~\cite{Buchmuller:1985jz} when writing Eq.~(\ref{ec:Fbuch}).}
The quark equations of motion can then be used in the last terms in Eqs.~(\ref{ec:OOdual}). Moreover, using the Bianchi identities it can be easily seen that
\begin{align}
& O_{q\tilde W}^{ij}  - (O_{q \tilde W}^{ji})^\dagger = 
  O_{q \tilde B}^{ij} - (O_{q \tilde B}^{ji})^\dagger = 
  O_{u \tilde B}^{ij} - (O_{u \tilde B}^{ji})^\dagger = 0 \,, \notag \\
& O_{q \tilde G}^{ij} - (O_{q \tilde G}^{ji})^\dagger = 
  O_{u \tilde G}^{ij} - (O_{u \tilde G}^{ji})^\dagger = 0 \,,
\end{align}
up to total derivatives.
Joining the two terms it is found that
\begin{small}
\begin{eqnarray}
O_{qW}^{ij} & = & -\frac{1}{4} \left[Y^u_{jk} \, O_{uW}^{ik} + Y^d_{jk} \, O_{dW}^{ik} 
  - Y^{u\dagger}_{ki} \, (O_{uW}^{jk})^\dagger - Y^{d\dagger}_{ki} \, (O_{dW}^{jk})^\dagger \right]
+ \frac{g}{4} \left[O_{\phi q}^{(3,ij)} + {O_{\phi q}^{(3,ji)}}^\dagger \right] \notag \\
& & + \frac{g}{4}  (\bar q_{Li} \gm \tau^I q_{Lj}) (\bar \ell_{Lk} \gamma_\mu \tau^I \ell_{Lk})
+ \frac{g}{4} (\bar q_{Li} \gm q_{Lj}) (\bar q_{Lk} \gamma_\mu \tau^I q_{Lk})
\,, \notag \\
O_{qB}^{ij} & = & -\frac{1}{4} \left[Y^u_{jk} \, O_{uB\phi}^{ik} + Y^d_{jk} \, O_{dB\phi}^{ik} - Y^{u\dagger}_{ki} \, (O_{uB\phi}^{jk})^\dagger - Y^{d\dagger}_{ki} \, (O_{dB\phi}^{jk})^\dagger \right]
+ \frac{g'}{4} \left[O_{\phi q}^{(1,ij)} + {O_{\phi q}^{(1,ji)}}^\dagger \right] \notag \\
& & - \frac{g'}{4} (\bar q_{Li} \gm q_{Lj}) (\bar \ell_{Lk} \gamma_\mu \ell_{Lk})
- \frac{g'}{2} (\bar q_{Li} \gm q_{Lj}) (\bar e_{Rk} \gamma_\mu e_{Rk})
+ \frac{g'}{12} (\bar q_{Li} \gm q_{Lj}) (\bar q_{Lk} \gamma_\mu q_{Lk}) \notag \\
& & + \frac{g'}{3} (\bar q_{Li} \gm q_{Lj}) (\bar u_{Rk} \gamma_\mu u_{Rk})
- \frac{g'}{6} (\bar q_{Li} \gm q_{Lj}) (\bar d_{Rk} \gamma_\mu d_{Rk}) \,, \notag \\
O_{uB}^{ij} & = & \frac{1}{4} \left[Y^u_{ki} \, O_{uB\phi}^{kj} - Y^{u\dagger}_{jk} \, (O_{uB\phi}^{ki})^\dagger \right]
+ \frac{g'}{4} \left[O_{\phi u}^{ij} + (O_{\phi u}^{ji})^\dagger \right] 
 - \frac{g'}{4} (\bar u_{Ri} \gm u_{Rj}) (\bar \ell_{Lk} \gamma_\mu \ell_{Lk}) \notag \\
& & 
- \frac{g'}{2} (\bar u_{Ri} \gm u_{Rj}) (\bar e_{Rk} \gamma_\mu e_{Rk})
+ \frac{g'}{12} (\bar u_{Ri} \gm u_{Rj}) (\bar q_{Lk} \gamma_\mu q_{Lk}) \notag \\
& & + \frac{g'}{3} (\bar u_{Ri} \gm u_{Rj}) (\bar u_{Rk} \gamma_\mu u_{Rk})
- \frac{g'}{6} (\bar u_{Ri} \gm u_{Rj}) (\bar d_{Rk} \gamma_\mu d_{Rk}) \,, \notag \\
O_{qG}^{ij} & = & -\frac{1}{4} \left[Y^u_{jk} \, O_{uG\phi}^{ik} + Y^d_{jk} \, O_{dG\phi}^{ik} - Y^{u\dagger}_{ki} \, (O_{uG\phi}^{jk})^\dagger - Y^{d\dagger}_{ki} \, (O_{dG\phi}^{jk})^\dagger \right]
 \notag \\[1mm]
& & + \frac{g_s}{4} (\bar q_{Li} \la \gm q_{Lj}) (\bar q_{Lk} \la \gamma_\mu q_{Lk})
+  \frac{g_s}{4} (\bar q_{Li} \la \gm q_{Lj}) (\bar u_{Rk} \la \gamma_\mu u_{Rk})  \notag \\[1mm]
& & +  \frac{g_s}{4} (\bar q_{Li} \la \gm q_{Lj}) (\bar d_{Rk} \la \gamma_\mu d_{Rk}) 
 \,, \notag \\
O_{uG}^{ij} & = & \frac{1}{4} \left[Y^u_{ki} \, O_{uG\phi}^{kj} - Y^{u\dagger}_{jk} \, (O_{uG\phi}^{ki})^\dagger \right] 
 + \frac{g_s}{4} (\bar u_{Ri} \la \gm u_{Rj}) (\bar q_{Lk} \la \gamma_\mu q_{Lk})
\notag \\[1mm]
& & +  \frac{g_s}{4} (\bar u_{Ri} \la \gm u_{Rj})(\bar u_{Rk} \la \gamma_\mu u_{Rk})
 +  \frac{g_s}{4} (\bar u_{Ri} \la \gm u_{Rj}) (\bar d_{Rk} \la \gamma_\mu d_{Rk}) \,.
\label{ec:Oredrew}
\end{eqnarray}
\end{small}%
A sum over $k=1,2,3$ is understood. The operators
\begin{align}
& O_{dB\phi}^{ij} = (\bar q_{Li} \smn d_{Rj}) \phi \Bmn \,, \notag \\
& O_{dG\phi}^{ij} = (\bar q_{Li} \la \smn d_{Rj}) \phi \Gmna
\end{align}
appearing in the above equations do not contribute to top couplings.

\subsection{Equalities for $O_{Du}^{ij}$, $O_{\bar Du}^{ij}$, $O_{Dd}^{ij}$ and $O_{\bar Dd}^{ij}$}

In order to show that these operators are redundant, it is convenient to consider their sums
$O_{Du}^{ij} + O_{\bar Du}^{ij}$, $O_{Dd}^{ij} + O_{\bar Dd}^{ij}$ and differences
$O_{Du}^{ij} - O_{\bar Du}^{ij}$, $O_{Dd}^{ij} - O_{\bar Dd}^{ij}$.
The sums can be written as
\begin{eqnarray}
O_{Du}^{ij} + O_{\bar Du}^{ij} & = & D_\mu (\bar q_{Li} u_{Rj}) D^\mu \tilde \phi
= - \bar q_{Li} u_{Rj} D_\mu D^\mu \tilde \phi \,, \notag \\
O_{Dd}^{ij} + O_{\bar Dd}^{ij} & = & D_\mu (\bar q_{Li} d_{Rj}) D^\mu \phi
= - \bar q_{Li} d_{Rj} D_\mu D^\mu \phi \,.
\end{eqnarray}
Using the scalar equations of motion it is found that these sums are equivalent to
\begin{eqnarray}
O_{Du}^{ij} + O_{\bar Du}^{ij} & = & - m^2 \bar q_{Li} u_{Rj} \tilde \phi + \lambda O_{u\phi}^{ij}
+ Y^e_{kl} (\bar q_{Li} u_{Rj}) \left[ (\bar \ell_{Lk} \epsilon)^T e_{Rl} \right] \notag \\
& & + Y^{u\dagger}_{kl} (\bar q_{Li} u_{Rj}) (\bar u_{Rk} q_{Ll}) 
 + Y^d_{kl} (\bar q_{Li} u_{Rj}) \left[ (\bar q_{Lk} \epsilon)^T d_{Rl} \right] \,, \notag \\
O_{Dd}^{ij} + O_{\bar Dd}^{ij} & = & - m^2 \bar q_{Li} d_{Rj} \phi + \lambda O_{d\phi}^{ij} 
+ Y^{e\dagger}_{kl} (\bar q_{Li} d_{Rj}) (\bar e_{Rk} \ell_{Ll}) \,, \notag \\
& & + Y^u_{kl} (\bar q_{Li} d_{Rj}) \left[ (\bar q_{Lk} \epsilon)^T u_{Rl} \right] 
+ Y^{d\dagger}_{kl} (\bar q_{Li} d_{Rj}) (\bar d_{Rk} q_{Ll}) \,,
\label{ec:rewOD1}
\end{eqnarray}
summing over $k,l$, with
\begin{eqnarray}
O_{u\phi}^{ij} & = & (\phi^\dagger \phi) \bar q_{Li} u_{Rj} \tilde \phi \,, \notag \\
O_{d\phi}^{ij} & = & (\phi^\dagger \phi) \bar q_{Li} d_{Rj} \phi \,,
\end{eqnarray}
and therefore redundant.
Note that the rewritten terms on the right-hand side of these equations do not contribute to the gauge boson vertices. This result is not surprising since the contribution before rewriting is proportional to $q^\mu$ and vanishes if the gauge boson is on-shell or coupling to masless external fermions.

In order to rewrite the differences $O_{Du}^{ij} - O_{\bar Du}^{ij}$ and $O_{Dd}^{ij} - O_{\bar Dd}^{ij}$, we introduce the auxiliary operators
\begin{align}
& {O}^{'ij}_{Du} = i (\bar q_{Li} \, \smn  D_\nu u_{Rj}) \, D_\mu \,\tilde \phi
\,, \notag \\
& {O}^{'ij}_{\bar D u} = i (D_\nu \bar q_{Li} \, \smn  u_{Rj}) \, D_\mu \,\tilde \phi \,, \notag \\
& {O}^{'ij}_{Dd} = i (\bar q_{Li} \, \smn  D_\nu d_{Rj}) \, D_\mu \, \phi
\,, \notag \\
& {O}^{'ij}_{\bar D d} = i (D_\nu \bar q_{Li} \, \smn  d_{Rj}) \, D_\mu \, \phi \,,
\end{align}
which are gauge invariant.
Using the definition of the $\smn$ matrices and $\{ \gm,\gn \}=2 \gmn$, it is easy to see that these two sets of operators are related by
\begin{align}
& O_{Du}^{ij} = {O}^{'ij}_{Du} - i \bar q_{Li} \gm \left( i \Dsl u_{Rj} \right) \, D_\mu \,\tilde \phi
&& \equiv {O}^{'ij}_{Du} - \Delta O^{ij}_{Du} 
\,, \notag \\
& O_{\bar D u}^{ij} = - {O}^{'ij}_{\bar D u}  + i \overline{\left( i \Dsl q_{Li} \right)} \gm u_{Rj} \, D_\mu \,\tilde \phi
&& \equiv - {O}^{'ij}_{\bar D u} + \Delta O^{ij}_{\bar D u} 
\,, \notag \\
& O_{Dd}^{ij} = {O}^{'ij}_{Dd} - i \bar q_{Li} \gm \left( i \Dsl d_{Rj} \right) \, D_\mu \, \phi 
&& \equiv {O}^{'ij}_{Dd} - \Delta O^{ij}_{Dd}
 \,, \notag \\
& O_{\bar D d}^{ij} = - {O}^{'ij}_{\bar D d}  + i \overline{\left( i \Dsl q_{Li} \right)} \gm d_{Rj} \, D_\mu \, \phi 
&& \equiv - {O}^{'ij}_{\bar D d} + \Delta O^{ij}_{\bar D d}
\,.
\label{ec:relO}
\end{align}
Using the fermion equations of motion we can obtain after a little algebra that
\begin{align}
& \Delta O^{ij}_{Du} = \frac{1}{2} Y^{u\dagger}_{jk} \left[ (O_{\phi q}^{(3,ki)})^\dagger - (O_{\phi q}^{(1,ki)})^\dagger \right] \,, \notag \\
& \Delta O^{ij}_{\bar Du} = - Y_{ki}^{u\dagger} (O_{\phi u}^{jk})^\dagger + Y_{ki}^{d\dagger} (O_{\phi \phi}^{jk})^\dagger \,, \notag \\
& \Delta O^{ij}_{Dd} = \frac{1}{2} Y^{d\dagger}_{jk} \left[ O_{\phi q}^{(3,ik)} + O_{\phi q}^{(1,ik)} \right] \,, \notag \\
& \Delta O^{ij}_{\bar Dd} = Y_{ki}^{u\dagger} O_{\phi \phi}^{kj} + Y_{ki}^{d\dagger} O_{\phi d}^{kj} \,,
\label{ec:OdeltaR}
\end{align} 
with
\begin{equation}
O_{\phi d}^{ij} = i (\phi^\dagger D_\mu \phi) (\bar d_{Ri} \gm d_{Rj}) \,.
\end{equation}
On the other hand, we have
\begin{eqnarray}
O_{Du}^{'ij} + O_{\bar Du}^{'ij} & = & i D_\nu (\bar q_{Li} \smn u_{Rj}) D_\mu \tilde \phi
= - i \bar q_{Li} \smn u_{Rj} D_\nu D_\mu \tilde \phi \,, \notag \\
O_{Dd}^{'ij} + O_{\bar Dd}^{'ij} & = & i D_\nu (\bar q_{Li} \smn d_{Rj}) D_\mu \phi
= - i \bar q_{Li} \smn d_{Rj} D_\nu D_\mu \phi \,.
\end{eqnarray}
Since
\begin{eqnarray}
[D_\mu,D_\nu] \phi & = & i g \frac{\tau^I}{2} \Wmn^I \phi + i g' Y \Bmn \phi  \,,
\end{eqnarray}
these sums can also be written in terms of already known operators,
\begin{eqnarray}
O_{Du}^{'ij} + O_{\bar Du}^{'ij} & = & -\frac{g}{4} O_{uW}^{ij} + \frac{g'}{4} O_{uB\phi}^{ij}
\,, \notag \\
O_{Dd}^{'ij} + O_{\bar Dd}^{'ij} & = & - \frac{g}{4} O_{dW}^{ij} - \frac{g'}{4} O_{dB\phi} \,.
\label{ec:OpR}
\end{eqnarray}
Finally, using Eqs.~(\ref{ec:relO}), (\ref{ec:OdeltaR}) and (\ref{ec:OpR}) we arrive at the desired result,
\begin{eqnarray}
O_{Du}^{ij} - O_{\bar Du}^{ij} & = & -\frac{g}{4} O_{uW}^{ij} + \frac{g'}{4} O_{uB\phi}^{ij}
- \frac{1}{2} Y^{u\dagger}_{jk} \left[ (O_{\phi q}^{(3,ki)})^\dagger - (O_{\phi q}^{(1,ki)})^\dagger \right] \notag \\
& & + Y_{ki}^{u\dagger} (O_{\phi u}^{jk})^\dagger - Y_{ki}^{d\dagger} (O_{\phi \phi}^{jk})^\dagger
\,, \notag \\
O_{Dd}^{ij} - O_{\bar Dd}^{ij} & = & - \frac{g}{4} O_{dW}^{ij} - \frac{g'}{4} O_{dB\phi}
- \frac{1}{2} Y^{d\dagger}_{jk} \left[ O_{\phi q}^{(3,ik)} + O_{\phi q}^{(1,ik)} \right] \notag \\
& & - Y_{ki}^{u\dagger} O_{\phi \phi}^{kj} - Y_{ki}^{d\dagger} O_{\phi d}^{kj} \,.
\end{eqnarray}
This, together with Eqs.~(\ref{ec:rewOD1}), shows that the four operators $O_{Du}^{ij}$, $O_{\bar Du}^{ij}$, $O_{Dd}^{ij}$ and $O_{\bar Dd}^{ij}$ are redundant.

\newpage
\section{General $Wtb$, $Ztt$, $\gamma tt$ and $gtt$ interactions}
\label{sec:3}

We consider new physics contributions to the third generation interactions, which are described by the dimension-six effective operators in Eqs.~(\ref{ec:Oall}) and (\ref{ec:Ostr}) with flavour indices $i=j=3$. These contributions are collected in appendix~\ref{sec:a} for complex coefficients $C_x$. Here we exclude from the analysis the redundant operators in these equations, in the same way as many other possible redundant operators one may construct~\cite{Buchmuller:1985jz} are ignored.
Thus, we provide completely general expressions for the $Wtb$, $Ztt$, $\gamma tt$ and $gtt$ vertices for off-shell fermions and bosons which also involve a minimal set of couplings. We remark that
the expressions presented here do not make any assumption about quark masses and mixings. (In fact, the operator equalities in the previous section involve arbitrary Yukawa matrices, with arbitrary masses and mixings.)
Moreover, we take into account all contributing operators independently of whether they also give corrections to the $Zbb$ verte, which is very constrained by present experimental data, or not. Cancellations among effective operator contributions are possible and occur in minimal SM extensions~\cite{delAguila:2000rc} as we will show later in more detail. (The same remarks apply to the results in the next section.) We finally show that anomalous coupling measurements might be used to determine effective operator coefficients.

\subsection{$Wtb$ vertex}

The effective $Wtb$ vertex including SM contributions and those from dimension-six operators can be parameterised as
\begin{eqnarray}
\mathcal{L}_{Wtb} & = & - \frac{g}{\sqrt 2} \bar b \, \gamma^{\mu} \left( \vl P_L
  + \vr P_R \right) t\; W_\mu^- 
   \nonumber \\
& & - \frac{g}{\sqrt 2} \bar b \, \frac{i \sigma^{\mu \nu} q_\nu}{M_W}
  \left( \gl P_L + \gr P_R \right) t\; W_\mu^- + \mathrm{H.c.} \,.
\label{ec:Wtb}
\end{eqnarray}
The mass scale normalising the $\smn q_\nu$ term has been taken as $M_W$ because this choice considerably simplifies the algebraic expressions of observables calculated from this vertex~\cite{AguilarSaavedra:2006fy}. Additionally, with this normalisation the relation between $\gl$, $\gr$ and effective operator coefficients is simpler and involves the ratio of scales $v^2/\Lambda^2$.
Within the SM, $\vl$ equals the Cabibbo-Kobayaski-Maskawa matrix element $V_{tb}\simeq 1$, while the rest of couplings
$\vr$, $\gl$ and $\gr$ vanish at the tree level. The contributions to these couplings from the operators in Eqs.~(\ref{ec:Oall}) are\footnote{Notice a typo in Eqs.~(8) of Ref.~\cite{AguilarSaavedra:2008gt}, where a minus sign should multiply $O_{uW}$.}
\begin{align}
& \delta \vl = C_{\phi q}^{(3,33)*}
     \frac{v^2}{\Lambda^2}   \,, 
&& \delta \gl = \sqrt 2 C_{dW}^{33*} \frac{v^2}{\Lambda^2} 
  \,, \notag \\
&  \delta \vr = \frac{1}{2} C_{\phi \phi}^{33} \frac{v^2}{\Lambda^2} \,,
&& \delta \gr = \sqrt 2 C_{uW}^{33} \frac{v^2}{\Lambda^2}  \,.
\end{align}
After removing redundant operators the structure of the Lagrangian in Eq.~(\ref{ec:Wtb}) is rather simple. All the new physics effects on the $Wtb$ vertex can be described by four parameters, which have a direct connection with effective operator coefficients.

\subsection{$Ztt$ vertex}
\label{sec:3.2}

We parameterise the $Ztt$ vertex including the SM contributions as well as those from dimension-six effective operators as
\begin{eqnarray}
\mathcal{L}_{Ztt} & = & - \frac{g}{2 c_W} \bar t \, \gm \left( \xttl P_L
  + \xttr P_R - 2 s_W^2 Q_t \right) t\; Z_\mu  \nonumber \\
& & - \frac{g}{2 c_W} \bar t \, \frac{i \smn q_\nu}{M_Z}
  \left( \dvz + i \daz \gamma_5 \right) t\; Z_\mu  \,,
\label{ec:Ztt}
\end{eqnarray}
with $Q_t=2/3$ the top quark electric charge. The mass scale for the $\smn q_\nu$ term is taken as $M_Z$ in analogy with the $Wtb$ vertex but, on the other hand, we have parameterised this coupling in terms of the vector and axial parts. The former is real while the latter is purely imaginary and CP-violating. They are the weak analogous to the top quark magnetic and electric dipole moment, respectively (see next subsection), up to normalisation.
Within the SM, these couplings take the values $\xttl = 2 \, T_3(t_L) = 1$, $\xttr = 2 \, T_3 (t_R) = 0$, where $T_3$ denotes the third isospin component, and $\dvz = \daz = 0$ at the tree level. The contributions from dimension-six operators are
\begin{align}
& \delta \xttl = \RE \left[  C_{\phi q}^{(3,33)} - C_{\phi q}^{(1,33)}
\right]  \frac{v^2}{\Lambda^2} \,,
&& \delta \dvz = \sqrt 2 \, \RE \left[ c_W  C_{uW}^{33} - s_W C_{uB\phi}^{33} \right]  \frac{v^2}{\Lambda^2}
\,, \notag \\
& \delta \xttr = - \RE \, C_{\phi u}^{33} \frac{v^2}{\Lambda^2} \,,
&& \delta \daz = \sqrt 2 \, \IM \left[ c_W C_{uW}^{33} - s_W  C_{uB\phi}^{33} \right] \frac{v^2}{\Lambda^2}
\,.
\end{align}
At this point it is worthwhile to discuss the relation between the $Ztt$ and $Zbb$ vertices, the latter very constrained by LEP data. Some authors drop from their analyses the operators $O_{\phi q}^{(3)}$ and $O_{\phi q}^{(1)}$ for this reason, because they give contributions
\begin{equation}
\delta X_{bb}^L = \RE \left[  C_{\phi q}^{(3,33)} + C_{\phi q}^{(1,33)} \right]
    \frac{v^2}{\Lambda^2} \,.
\end{equation}
However, cancellations are possible and take place in minimal models. For example, in a SM extension with a $Q=2/3$ singlet we have $C_{\phi q}^{(3,33)} =
- C_{\phi q}^{(1,33)}$~\cite{delAguila:2000rc}, so that the contribution to the $Zbb$ vertex identically cancels but there can be deviations in the $Ztt$ interaction.
One can still wonder about the corrections to the $Wtb$ vertex from $O_{\phi q}^{(3,33)}$, which may affect low-energy $B$ physics~\cite{Fox:2007in}. However, in this particular model additional contributions can (at least partly) make up for the difference as it has been shown with a global analysis of precision electroweak data and low energy constraints from $B$ and $K$ physics~\cite{AguilarSaavedra:2002kr}. The reason behind the (partial) cancellation among new physics contributions in this simple, particular model is precisely the Glashow-Iliopoulos-Maiani (GIM) mechanism~\cite{Glashow:1970gm}.

\subsection{$\gamma tt$ vertex}

The $\gamma tt$ vertex including the SM coupling (given by the top electric charge $Q_t$) and contributions from dimension-six effective operators can be parameterised as
\begin{equation}
\mathcal{L}_{\gamma tt} = - e Q_t \bar t \, \gm  t\; A_\mu  - e \bar t \, \frac{i \smn q_\nu}{m_t} \left( \dva + i \daa \gamma_5 \right) t\; A_\mu  \,.
\label{ec:gatt}
\end{equation}
The couplings $\dva$, $\daa$ are real and related to the top quark magnetic and electric dipole moment, respectively, by a multiplicative factor, and the latter is CP-violating.
For this interaction we have
\begin{eqnarray}
\delta \dva & = & \frac{\sqrt 2}{e} \, \RE \left[ c_W C_{uB\phi}^{33} + s_W  C_{uW}^{33}\right] \frac{v m_t}{\Lambda^2}
  \,, \notag \\
\delta \daa & = & \frac{\sqrt 2}{e} \, \IM \left[ c_W C_{uB\phi}^{33} + s_W  C_{uW}^{33} \right] \frac{v m_t}{\Lambda^2} \,.
\end{eqnarray}
We note that the $\gm$ term does not receive corrections from dimension-six operators (this also applies to the $gtt$, $\gamma tu/\gamma tc$ and $gtu/gtc$ vertices). If we had included the redundant operators $O_{qW}$, $O_{qB}$ and $O_{uB}$, the first two would yield corrections $\sim q^2 \bar t_L \gm t_L A_\mu$ and the latter $\sim q^2 \bar t_R \gm t_R A_\mu$, non-vanishing only when the photon is off-shell. The operator rewriting in Eqs.~(\ref{ec:Oredrew}) eliminates such terms, so that corrections to the electromagnetic coupling are absent even for off-shell photons. In particular, as dictated by Eqs.~(\ref{ec:Oredrew})
the contribution to the amplitudes of a $q^2$-dependent $\gm$ term
can be reproduced by a constant $\gm$ term (proportional to the square of the gauge boson mass) plus four-fermion interactions. An explicit example can be found in appendix~\ref{sec:c}.

\subsection{$gtt$ vertex}

This vertex, including the SM contribution, is written as
\begin{equation}
\mathcal{L}_{gtt} = - g_s \bar t \, \frac{\la}{2} \gm  t\; G_\mu^a  - g_s \bar t \, \la \frac{i \smn q_\nu}{m_t} \left( \dvG + i \daG \gamma_5 \right) t\; G_\mu^a  \,,
\label{ec:gtt}
\end{equation}
The couplings $\dvG$, $\daG$ are real and related to the top chromomagnetic and chromoelectric dipole moments, respectively. They vanish in the SM at the tree level. The new physics contributions from effective operators are
\begin{eqnarray}
\delta \dvG & = & \frac{\sqrt 2}{g_s} \RE C_{uG\phi}^{33} \frac{v m_t}{\Lambda^2}
  \,, \notag \\
\delta \daG & = & \frac{\sqrt 2}{g_s} \IM C_{uG\phi}^{33} \frac{v m_t}{\Lambda^2} \,.
\end{eqnarray}
As in the case of the photon, the $\gm$ term does not include corrections from dimension-six operators.

\subsection{Determination of effective operator coefficients}

One may finally wonder whether hypothetical measurements of these anomalous couplings might provide any insight on the effective operators generating them. 
The answer is affirmative, since there are 12 anomalous couplings and only 8 operator coefficients. (Notice, however, that some of the anomalous couplings correspond to the real or imaginary parts of an effective operator coefficient or combination of them.)
The measurement of $Wtb$ anomalous couplings would translate into a measurement of $C_{\phi q}^{3,33}$, $C_{\phi \phi}^{33}$, $C_{dW}^{33}$ and $C_{uW}^{33}$. In the $Ztt$ vertex, the anomalous contributions to $\xttl$ and $\xttr$ would then determine the real parts of $C_{\phi q}^{(1,33)}$ and $C_{\phi q}^{33}$.
From $\dvz$ and $\dva$ the real parts of $C_{uW}^{33}$ and $C_{uB\phi}^{33}$ might be obtained, while from $\daz$ and $\daa$ one would obtain the imaginary parts.
In the gluon vertex, $\dvG$ and $\daG$ determine the real and imaginary parts, respectively, of $C_{uG\phi}^{33}$.
The only coefficient for which two independent determinations are possible is $C_{uW}^{33}$, which could be obtained from $\gr$ and also from the combined measurements of $\dvz$, $\dva$ (the real part) and $\daz$, $\daa$ (the imaginary part).
Of course, this is a tremendously optimistic picture, since obtaining these measurements in a real detector is very challenging (see for instance Ref.~\cite{AguilarSaavedra:2007rs}) and finding some evidence of physics beyond the SM would already be very positive.

\section{General top flavour-changing interactions}
\label{sec:4}

In this section we collect the general Lagrangians for $Ztc$, $\gamma tc$ and $gtc$ interactions with off-shell $t$, $c$ quarks and gauge bosons, and the relation between the respective terms and coefficients of dimension-six gauge-invariant operators. For $Ztu$, $\gamma tu$ and $gtu$ vertices the Lagrangian structure is the same and the coefficients are obtained by replacing the generation index ($2 \to 1$). The contributions of all operators are collected in appendix~\ref{sec:b}. 

\subsection{$Ztc$ vertex}

This interaction, as the remaining flavour-changing ones, vanishes in the SM at the tree level due to the GIM mechanism. The contributions from dimension-six operators can be
parameterised with the Lagrangian
\begin{eqnarray}
\mathcal{L}_{Ztc} & = & - \frac{g}{2 c_W} \bar c \, \gm \left( \xctl P_L
  + \xctr P_R \right) t\; Z_\mu  \nonumber \\
& & - \frac{g}{2 c_W} \bar c \, \frac{i \smn q_\nu}{M_Z}
  \left( \kl P_L + \kr P_R \right) t\; Z_\mu  + \text{H.c.}  \,,
\label{ec:Ztc}
\end{eqnarray}
including only $\gm$ and $\smn q_\nu$ terms and involving four anomalous couplings whose contributions from effective operators read
\begin{align}
& \delta \xctl = \frac{1}{2} \left[ C_{\phi q}^{(3,23)} + C_{\phi q}^{(3,32)*} 
      - C_{\phi q}^{(1,23)} - C_{\phi q}^{(1,32)*} \right] \frac{v^2}{\Lambda^2}
\,, \notag \\
& \delta \xctr = - \frac{1}{2} \left[ C_{\phi u}^{23} + C_{\phi u}^{32*} 
  \right] \frac{v^2}{\Lambda^2}
\,, \notag \\
& \delta \kl = \sqrt 2 \left[ c_W C_{uW}^{32*} - s_W C_{uB\phi}^{32*} 
  \right] \frac{v^2}{\Lambda^2}
\,, \notag \\
& \delta \kr = \sqrt 2 \left[ c_W C_{uW}^{23} - s_W C_{uB\phi}^{23} \right]
 \frac{v^2}{\Lambda^2}
\,.
\label{ec:Ztcex}
\end{align}
A few comments are now in order. The Lagrangian in Eq.~(\ref{ec:Ztc}) for $Ztc$ interactions involves all contributing dimension-six effective operators,
including $O_{\phi q}^{(3,ij)}$ and $O_{\phi q}^{(1,ij)}$ that were discarded
in Refs.~\cite{Ferreira:2008cj,Coimbra:2008qp} because cancellations were banned there.
These cancellations naturally happen in some SM extensions, for example with an extra $Q=2/3$ quark singlet (see the discussion in section~\ref{sec:3.2} and Refs.~\cite{delAguila:2000rc,AguilarSaavedra:2002kr}). Hence, this simplification does not have a solid phenomenological basis and turns out to be restrictive.
We also observe that, although there are
several possible contributing operators, the number of relevant parameters necessary to describe the $Ztc$ interaction is only four instead of seven as considered before~\cite{Ferreira:2008cj,Coimbra:2008qp}. We also emphasise that these parameters are independent, as we will observe by comparing with the results in the next subsection. We finally remark that a Lagrangian equivalent to the one in Eq.~(\ref{ec:Ztc}) has been used for previous phenomenological analyses~\cite{delAguila:1999ac,delAguila:1999ec}. The results presented here show that those analyses are completely general also if regarded from the framework of dimension-six gauge-invariant effective operators.

\subsection{$\gamma tc$ vertex}

The $\gamma tc$ vertex arising from dimension-six effective operators can be parameterised in full generality as
\begin{equation}
\mathcal{L}_{\gamma tc} = - e \bar c \, \frac{i \smn q_\nu}{m_t} \left( \lL P_L
+ \lR P_R \right) t\; A_\mu  + \text{H.c.}  \,,
\label{ec:gatc}
\end{equation}
where
\begin{eqnarray}
\delta \lL & = & \frac{\sqrt 2}{e} \left[ s_W C_{uW}^{32*} + c_W C_{uB\phi}^{32*}
   \right] \frac{v m_t}{\Lambda^2}
\,, \notag \\
\delta \lR & = & \frac{\sqrt 2}{e} \left[ s_W C_{uW}^{23} + c_W C_{uB\phi}^{23}
    \right] \frac{v m_t}{\Lambda^2}
\,.
\end{eqnarray}
This Lagrangian is completely general but only involves two independent parameters instead of the four ones considered in Refs.~\cite{Ferreira:2008cj,Coimbra:2008qp}. As we have anticipated, these parameters are independent from the corresponding ones involving a $Ztc$ interaction. (For example, a combination of operators with $C_{uB\phi}^{32} = -\tan \theta_W C_{uW}^{23}$ does not contribute to $\lR$ but contributes to $\kr$.) An equivalent Lagrangian has been used in previous phenomenological analyses~\cite{delAguila:1999ac,delAguila:1999ec} which are completely general, as we have shown here.

\subsection{$gtc$ vertex}

Finally, the $gtc$ vertex arising from dimension-six effective operators is written as
\begin{equation}
\mathcal{L}_{gtc} = - g_s \bar c \, \la \frac{i \smn q_\nu}{m_t} \left( \gcl P_L + \gcr P_R \right) t\; G_\mu^a  \,,
\label{ec:gct}
\end{equation}
where the contributions to the two relevant couplings $\gcl$, $\gcr$ are
\begin{eqnarray}
\delta \gcl & = & \frac{\sqrt 2}{g_s} C_{uG\phi}^{32*} \frac{v m_t}{\Lambda^2}
\,, \notag \\
\delta \gcr & = & \frac{\sqrt 2}{g_s} C_{uG\phi}^{23} \frac{v m_t}{\Lambda^2}
\,.
\end{eqnarray}
This parameterisation seems more convenient that other ones \cite{Ferreira:2005dr} because the rewriting of $O_{qG}$ and $O_{uG}$ eliminates quartic terms which otherwise would have to be included in amplitude calculations. Of course, the physical results are independent of the parameterisation but the effort needed for computations may be reduced if an adequate parameter set is chosen.

\subsection{Determination of effective operator coefficients}

A hypothetical measurement of FCN couplings might eventually be used to determine effective operator coefficients but not completely, because there are 8 anomalous couplings involved in the vertices and 9 effective operator coefficients. Let us focus on top-charm interactions for definiteness. The simultaneous measurement of $\kl$ and $\lL$ would determine $C_{uW}^{32}$ and $C_{uB\phi}^{32}$, while $\kr$ and $\lR$ would do the same for $C_{uW}^{23}$ and $C_{uB\phi}^{23}$. From the gluon couplings, $C_{uG\phi}^{32}$ and $C_{uG\phi}^{23}$ might be obtained. On the other hand, the various coefficients appearing in $\xctl$ and $\xctr$ cannot be disentangled using only measurements of FCN couplings. As we have pointed out, the FCN anomalous couplings are all independent, although they could be related to anomalous couplings appearing in the $Wtd$ and $Wts$ vertices, which are not addressed here.

\section{Summary}

New physics at a higher scale can be described within the framework of gauge-invariant effective operators which result after integrating the heavy degrees of freedom. These  operators can induce corrections to SM couplings and, in particular, may originate anomalous couplings of the top quark to the gauge bosons. The large number and variety of dimension-six gauge-invariant effective operators~\cite{Buchmuller:1985jz} leads to the apperance of many possible Lorentz structures for the top trilinear vertices involving a large number of parameters.

In this work we have used the equations of motion to remove redundant operators,
arriving at the gratifying conclusion that {\em all} effective operator contributions to the trilinear $V \Fi \Fj$ vertices involving a $W$ or $Z$ boson, a photon or a gluon, can be parameterised in full generality using only $\gm$ and $\smn q_\nu$ terms, with $q=p_j-p_i$.
This result, which is well-known for an on-shell boson $V$ and on-shell fermions $\Fi$, $\Fj$, is also valid if they are off-shell due to the gauge structure of the theory, {\em i.e.} the fact that gauge-invariant effective operators include not only $V\Fi\Fj$ vertices but also other ones as for example $gV\Fi\Fj$ and four-fermion interactions. In this way, phenomenological analyses involving anomalous couplings can be considerably simplified. Compared to previous literature, we find that new physics contributions to top interactions can be described with a smaller number of parameters and a simpler Lorentz structure. The reduction in the number of effective operator coefficients can be read in Table~\ref{tab:summ} for each of the couplings studied. The number of anomalous couplings involved in the vertices is included as well. Note that some of the anomalous couplings are real or purely imaginary by definition. For example, in the $gtt$ vertex the two anomalous couplings involve the real and imaginary part of the coefficient $C_{uG\phi}^{33}$. 

\begin{table}[htb]
\begin{center}
\begin{tabular}{lccccccc}
                & $Wtb$ & $Ztt$ & $\gamma tt$ & $gtt$ & $Ztu/c$ & $\gamma tu/c$ & $gtu/c$  \\
$C_x^0$           & 9     & 10    & 5           & 3     & 20      & 10            & 6  \\
$C_x$             & 4     & 5     & 2           & 1     & 10      & 4             & 2  \\
$N_\text{min}$    & 4     & 4     & 2           & 2     & 4       & 2             & 2
\end{tabular}
\caption{For each interaction: number of effective operator coefficients $C_x^0$, $C_x$ contributing to the trilinear vertex before and after removing redundant operators; number of anomalous couplings $N_\text{min}$ necessary to describe the vertex.}
\label{tab:summ}
\end{center}
\end{table}

A second important point which must be noted is that in most cases the relation between anomalous couplings and effective operator coefficients is direct, for example in the $Wtb$ vertex.
This would allow for the determination of effective operator coefficients if these anomalous couplings were measured at LHC~\cite{AguilarSaavedra:2008gt}. In some other cases effective operator coefficients can be determined by simultaneous measurements of two anomalous couplings, {\em e.g.} involving the $Z$ boson and the photon couplings.
Moreover, if all anomalous couplings in the $Wtb$, $Ztt$ and $\gamma tt$ vertices might be determined (which is an extremely optimistic assumption) a consistency check could be performed by comparing the determination of $C_{uW}^{33}$ from the $Wtb$ coupling $\gr$ and from the simultaneous measurement of the $Ztt$ couplings $\dvz$, $\daz$ and $\gamma tt$ couplings $\dva$, $\daa$. Analogous tests are not possible in FCN interactions where all anomalous couplings involve independent combinations of operator coefficients, but could be possible if
anomalous $Wtd$ and $Wts$ interactions were included.

Working within the framework of gauge-invariant effective operators, phenomenological studies of the influence of top anomalous couplings can be carried out using the simple Lagrangians given in sections~\ref{sec:3} and \ref{sec:4}. (This is valid for any other fermion, since the results are general.) Of course, in a given process there may be further contributions to the amplitudes apart from those originating from trilinear vertices, related by gauge symmetry. But in this respect the operator rewriting performed also proves to be very useful. The redundant operators removed include associated interactions, for example $gZtc$ and $g\gamma tc$ vertices, which should otherwise be taken into account in amplitude calculations. After rewriting these operators and showing that they are redundant, not only the extra terms in the vertex are unnecessary but also the quartic interactions. This fact greatly simplifies the theoretical setup and the development of Monte Carlo generators involving top anomalous couplings.

\section*{Acknowledgements}

I thank M. P\'erez-Victoria for discussions and  F. del Aguila and J. Wudka for useful comments. This work has been supported by a MEC Ram\'on y Cajal contract, MEC project FPA2006-05294 and
Junta de Andaluc{\'\i}a projects FQM 101 and FQM 437.

\appendix

\section{Gauge symmetry and amplitude calculations}
\label{sec:c}

\begin{figure}[htb]
\begin{center}
\begin{tabular}{ccccc}
\epsfig{file=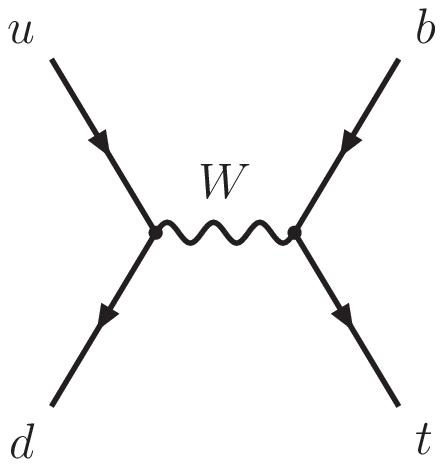,height=3cm,clip=} & \quad \quad &
\epsfig{file=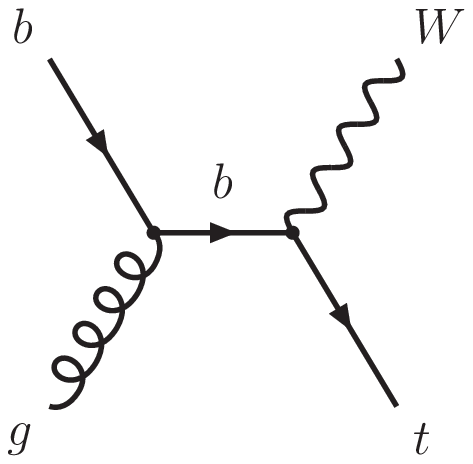,height=3cm,clip=} & \quad \quad &
\epsfig{file=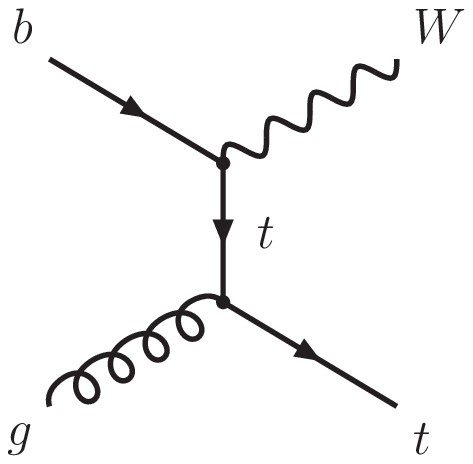,height=3cm,clip=}
\end{tabular}
\caption{Left: Feynman diagram for single top production in the
$u \bar d \to t \bar b$ process. Center, right: diagrams contributing to $gb \to tW^-$.}
\label{fig:diag-tbtW}
\end{center}
\end{figure}

For a better understanding of the results obtained, we show in this appendix how the relations among operators obtained actually translate into the rewriting of the vertices, and why this rewriting gives the same result in amplitude calculations. For simplicity we restrict ourselves to charged current $Wtb$ interactions and study the implications for two single top processes in hadron collisions: $s$-channel production $u \bar d \to t \bar b$ and $tW^-$ associated production, $gb \to tW^-$, depicted in Fig.~\ref{fig:diag-tbtW}.
Analogous results hold for the rest of operators and other processes, as it is expected, once that all contributions from gauge invariant operators are included. 

We begin with the operator $O_{qW}^{33}$, whose contribution to these two processes is given by the $Wtb$ and $gWtb$ vertices,
\begin{eqnarray}
\alpha O_{qW}^{33} + \alpha^* (O_{qW}^{33})^\dagger & \supset &
-\sqrt 2 \, \RE \alpha \;  q^2 \, \bar b_L \gm t_L \; \Wm^- \notag \\
& & + i \sqrt 2 \, \IM \alpha \left[ \bar b_L (\qsl k^\mu - k\cdot q \; \gm  ) t_L \right. \notag \\
& & \left.
+2 g_s \bar t_L \frac{\lambda^a}{2} (\gm p_W^\nu  - \pwsl g^{\mu \nu}) b_L G_\nu^a  \right] \Wm^- + \text{H.c.}
\label{ec:OqW}
\end{eqnarray}
where $q=p_t-p_b$ and $p_W$ is the outgoing $W$ boson momentum, which are equal in the triple vertex. On the other hand, using Eqs.~(\ref{ec:Oredrew}) to write $O_{qW}^{33}$ in terms of other operators, the relevant contributions are $Wtb$ vertices and a four-fermion interaction,
\begin{eqnarray}
\alpha O_{qW}^{33} + \alpha^* (O_{qW}^{33})^\dagger & \supset &
-\sqrt 2 \, \RE \alpha \left[M_W^2 \, \bar b_L \gm t_L \; \Wm^-  - \frac{g}{\sqrt 2} (\bar b_L \gm t_L) (\bar u_L \gamma_\mu d_L) + \dots \right] \notag \\
& &  i \sqrt 2 \, \IM \alpha \; \bar b \, i \smn q_\nu (m_t P_R - m_b P_L )
\; t \; \Wm^- + \text{H.c.}
\label{ec:OqW2}
\end{eqnarray}
This rewriting gives the same results in amplitude calculations even for off-shell fermions or bosons. In the $s$-channel process, the $t$ and $b$ quarks involved in the $Wtb$ vertex are on-shell but the $W$ boson is not. The non-trivial substitution is in this case
\begin{equation}
q^2 \, \bar b_L \gm t_L \; \Wm^- \to M_W^2 \, \bar b_L \gm t_L \; \Wm^-  - \frac{g}{\sqrt 2} (\bar b_L \gm t_L) (\bar u_L \gamma_\mu d_L)
\label{ec:sub1}
\end{equation}
As we will find with the explicit calculation below, the resulting amplitude is the same if we use (i) the $q^2 \bar b_L \gm t_L$ interaction on the left-hand side of this equation, or (ii) the $M_W^2 \bar b_L \gm t_L$ term on the right-hand side plus the four-fermion contribution.
In $tW^-$ associated production the $W$ boson is on-shell but the top and bottom quarks are not. For this process, the non-trivial substitution is
\begin{align}
& \left[ \bar b_L (\qsl k^\mu - k\cdot q \; \gm  ) t_L +2 g_s \bar t_L \frac{\lambda^a}{2} (\gm p_W^\nu  - \pwsl g^{\mu \nu}) b_L G_\nu^a  \right] \Wm^- \notag \\
& \to \bar b \, i \smn q_\nu (m_t P_R - m_b P_L ) \; t \; \Wm^- \,,
\label{ec:sub2}
\end{align}
which gives the same result for the $gb \to tW^-$ amplitude once that the contribution of the $Wtb$ vertex to both diagrams and the new diagram involving the quartic vertex, which is only present before the rewriting, are summed.
The same reckoning obviously applies to the $Z$ boson, the photon and the gluon,
namely the operators $O_{qB}^{ij}$, $O_{uB}^{ij}$, $O_{qG}^{ij}$ and
$O_{uG}^{ij}$. Note that for this latter case a different approach has been taken in Ref.~\cite{Ferreira:2005dr} and subsequent works, doing the opposite replacement
to Eq.~(\ref{ec:sub2}).
We find that performing the substitutions as suggested here has the added advantage of removing quartic interactions from the analysis.

For the rewriting of the combinations $O^{ij}_{Du}-O^{ij}_{\bar Du}$,
$O^{ij}_{Dd}-O^{ij}_{\bar Dd}$ the arguments are analogous.
The operator equalities imply for these processes the replacements
\begin{eqnarray}
\left[ \bar b_L k^\mu t_R - g_s \bar b_L \la \gmn t_R G_\nu^a \right] \Wm^-
& \to & \bar b_L i \smn q_\nu t_R  + m_t \bar b_L \gm t_L + m_b \bar b_R \gm t_R \,, \notag \\
\left[ \bar t_L k^\mu b_R - g_s \bar t_L \la \gmn b_R G_\nu^a \right] \Wm^+
& \to & - \bar t_L i \smn q_\nu b_R  + m_b \bar t_L \gm b_L + m_t \bar t_R \gm b_R \,,
\label{ec:subG1}
\end{eqnarray}
which give the same result in amplitude calculations \cite{AguilarSaavedra:2008gt}.
Notice that the trilinear term substitutions in these equations
exactly correspond to the Gordon identities that can be applied for on-shell fermions. In the off-shell case, {\em e.g.} in $tW^-$ production, the trilinear terms in both sides are not equal but their difference is compensated precisely by the $gWtb$ quartic vertex, which is not present on the right-hand side. Besides, it must be remarked that the rewritten expressions bring the advantage of not only removing the $k^\mu$ terms from the effective $Wtb$ vertex, but also the associated quartic $gWtb$ interactions which
otherwise should be included in some of the amplitudes.

In the following we carry out the amplitude calculations to check that the replacements in Eqs.~(\ref{ec:sub1}), (\ref{ec:sub2}) give the same results in $u \bar d \to t \bar b$ and $g b \to tW^-$. The same has been done in Ref.~\cite{AguilarSaavedra:2008gt} for the replacements in Eqs.~(\ref{ec:subG1}).

\subsection{Amplitude for $u \bar d \to t \bar b$}

We denote by $p_1$, $p_2$, $p_3$ and $p_4$ the momenta of the $u$, $\bar d$, $t$ and $\bar b$ quarks, respectively. Using the $Wtb$ interaction on the left-hand side of Eq.~(\ref{ec:sub1}) and the standard $Wud$ vertex, the amplitude reads
\begin{equation}
\mathcal{M}_1 = - g \frac{q^2}{q^2-M_W^2} \bar u(p_3) \gm P_L v(p_4) \; \bar v(p_2) \gamma_\mu P_L u(p_1) \,,
\end{equation}
while using the $Wtb$ interaction on the right-hand side the amplitude is
\begin{equation}
\mathcal{M}_2 = - g \frac{M_W^2}{q^2-M_W^2} \bar u(p_3) \gm P_L v(p_4) \; \bar v(p_2) \gamma_\mu P_L u(p_1) \,.
\end{equation}
The amplitude corresponding to the four-fermion interaction is
\begin{equation}
\mathcal{M}_3 = - g \bar u(p_3) \gm P_L v(p_4) \; \bar v(p_2) \gamma_\mu P_L u(p_1) \,,
\end{equation}
so that it is evident that $\mathcal{M}_1 = \mathcal{M}_2 + \mathcal{M}_3$, as it should be.

\subsection{Amplitude for $g b \to tW^-$}

Checking that the substitution in Eq.~(\ref{ec:sub2}) gives the same result in the $gb \to tW^-$ amplitude is algebraically much more involved. The computations can be considerably simplified if we define an ``off-shell''  operator subtracting the trilinear terms in this equation,
\begin{equation}
\mathcal{O}_3 = \left[ \bar b (\qsl k^\mu - k\cdot q \, \gm) P_L t -
\bar b i \smn q_\nu (m_t P_R - m_b P_L) t \right] W_\mu^- + \text{H.c.}
\end{equation}
Then, to prove the validity of substitution in Eq.~(\ref{ec:sub2}) we only have to show that the contribution of $\mathcal{O}_3$ plus the quartic term identically vanish.
Using the anticommutation relation for $\gamma$ matrices and the definition of $\smn$,
$\mathcal{O}_3$  can be written in a much more convenient form,
\begin{eqnarray}
\mathcal{O}_3 & = & \left[ \bar b \gm (m_t P_L - m_b P_R) (\ptsl - m_t) t -
\bar b (\pbsl - m_b) \gm (m_b P_L - m_t P_R) t \right] \Wm^- \notag \\[1mm]
& & + \left[ \bar b k^\mu P_R (\ptsl - m_t) t - \bar b (\pbsl - m_b) k^\mu P_L t
\right] \Wm^- \notag \\[1mm]
& & + \left[ - \bar b \gm P_L (p_t^2-m_t^2) t + \bar b (p_b^2-m_b^2) \gm P_L t \right] \Wm^- + \text{H.c.}
\label{ec:O3}
\end{eqnarray}
This expression makes it apparent that $\mathcal{O}_3$ vanishes for both $t$, $b$ on-shell.

We denote by $p_1$, $p_2$, $p_3$ and $p_4$ the momenta of the gluon, $b$, $t$ and $W^-$ boson, respectively. We use superscripts $a$, $b$, $c$ to label the contributions to the amplitudes of the three terms in Eq.~(\ref{ec:O3}), in the order shown, and subscripts $1,2$ corresponding to the $s$- and $t$-channel diagrams. After trivial simplifications
using $(\psl - m) (\psl + m) = p^2-m^2$, the first term gives
\begin{eqnarray}
\mathcal{M}_1^a & = & - \frac{g_s}{2} \bar u(p_3) \la \gm \gn (m_t P_L -m_b P_R) u(p_2) \times \varepsilon \,, \notag \\
\mathcal{M}_2^a & = & - \frac{g_s}{2} \bar u(p_3) \la \gn \gm (m_t P_L -m_b P_R) u(p_2) \times \varepsilon \,,
\end{eqnarray}
where $\varepsilon$ stands for the product of polarisation vectors $\varepsilon_\mu^* (p_4) \varepsilon_\nu(p_1)$. The sum of both diagrams is
\begin{equation}
\mathcal{M}_{1+2}^a = - g_s \bar u(p_3) \la \gmn (m_t P_L -m_b P_R) u(p_2) \times \varepsilon \,.
\end{equation}
The second term gives
\begin{eqnarray}
\mathcal{M}_1^b & = & \frac{g_s}{2} \bar u(p_3) \la (p_5+p_3)^\mu \gn P_L u(p_2)
\times \varepsilon \,, \notag \\
\mathcal{M}_2^b & = & \frac{g_s}{2} \bar u(p_3) \la (p_2+p_6)^\mu \gn P_L u(p_2)
\times \varepsilon \,,
\end{eqnarray}
where $p_5 = p_1+p_2$ and $p_6 = p_3-p_1$ are the momenta of the internal $b$, $t$ quarks in the $s$- and $t$-channel diagrams, respectively. The sum of both is
\begin{equation}
\mathcal{M}_{1+2}^b = g_s \bar u(p_3) \la p_1^\mu \gn P_L u(p_2) \times \varepsilon \,.
\end{equation}
The third term yields the contributions
\begin{eqnarray}
\mathcal{M}_1^c & = & - \frac{g_s}{2} \bar u(p_3) \la \left[
P_R \gm (p_2\!\!\!\!\!\!\!\not\,\,\,\,+m_b) \gn
+ \gm  p_1\!\!\!\!\!\!\!\not\,\,\,\, \gn P_L \right] u(p_2)
\times \varepsilon \,, \notag \\
\mathcal{M}_2^c & = & \frac{g_s}{2} \bar u(p_3) \la \left[
\gn (p_3\!\!\!\!\!\!\!\not\,\,\,\,+m_t) \gm P_L
- \gn p_1\!\!\!\!\!\!\!\not\,\,\,\, \gm P_L \right] u(p_2)
\times \varepsilon \,.
\end{eqnarray}
Using the $\gamma$ anticommutation relations and the Dirac equation for external fermions, the sum of both is
\begin{eqnarray}
\mathcal{M}_{1+2}^c & = & - g_s \bar u(p_3) \la p_1^\mu \gn P_L u(p_2) \times \varepsilon
- g_s \bar u(p_3) \la \gm p_4^\nu P_L u(p_2) \times \varepsilon \notag \\
& & + g_s \bar u(p_3) \la p_1\!\!\!\!\!\!\!\not\,\,\,\, \gmn P_L u(p_2) \times \varepsilon \,.
\end{eqnarray}
Notice that the first of these terms already cancels $\mathcal{M}_{1+2}^b$. Finally, the contribution of the quartic $gWtb$ coupling is
\begin{eqnarray}
\mathcal{M}_3 & = & - g_s \bar u(p_3) \la p_1\!\!\!\!\!\!\!\not\,\,\,\, \gmn P_L u(p_2) \times \varepsilon + g_s \bar u(p_3) \la (m_t P_L - m_b P_R) \gmn u(p_2) \times \varepsilon \notag \\
& & + g_s \bar u(p_3) \la \gm p_4^\nu P_L u(p_2) \times \varepsilon \,.
\end{eqnarray}
The sum
\begin{equation}
\mathcal{M}_{1+2}^a + \mathcal{M}_{1+2}^b + \mathcal{M}_{1+2}^c + \mathcal{M}_3 = 0
\end{equation}
vanishes, as expected.

\newpage
\section{Operator contributions to $Wtb$, $Ztt$, $\gamma tt$ and $gtt$}
\label{sec:a}

We collect here the contribution to the effective $Wtb$, $Ztt$, $\gamma tt$ and $gtt$ vertices of the operators in Eqs.~(\ref{ec:Oall}), (\ref{ec:Ostr}), including also those from the operators which are redundant. We also collect the contributions to
the associated $gWtb$, $gZtt$, $g\gamma tt$ and $ggtt$ quartic couplings. We use the shorthand $\alpha_x = C_x/\Lambda^2$ and drop the indices in the $\alpha$ constants.
Our expressions coincide with those in Refs.~\cite{Whisnant:1997qu,Yang:1997iv} except for sign differences originating from the different definitions of the covariant derivative and the $Z$ field, and also coincide with
Ref.~\cite{delAguila:2000aa}.
For the $Wtb$ interaction we have
\begin{eqnarray}
\alpha O_{\phi q}^{(3,33)} + \alpha^* (O_{\phi q}^{(3,33)})^\dagger & \supset &
 -\alpha \frac{gv^2}{\sqrt 2} \bar t_L \gm b_L \Wm^+
 -\alpha^* \frac{gv^2}{\sqrt 2} \bar b_L \gm t_L \Wm^-
  \,, \notag \\
\alpha O_{\phi \phi}^{33} + \alpha^* (O_{\phi \phi}^{33})^\dagger & \supset &
  -\alpha \frac{gv^2}{2\sqrt 2} \bar t_R \gm b_R \Wm^+
  - \alpha^* \frac{gv^2}{2\sqrt 2} \bar b_R \gm t_R \Wm^-
  \,, \notag \\
\alpha O_{uW}^{33} + \alpha^* (O_{uW}^{33})^\dagger & \supset &
  \alpha v \bar b_L \smn t_R \Wmn^- + \alpha^* v \bar t_R \smn b_L \Wmn^+
  \,, \notag \\
\alpha O_{dW}^{33} + \alpha^* (O_{dW}^{33})^\dagger & \supset &
  \alpha v \bar t_L \smn b_R \Wmn^+ + \alpha^* v \bar b_R \smn t_L \Wmn^-
  \,, \notag \\
\alpha O_{Du}^{33} + \alpha^* (O_{Du}^{33})^\dagger & \supset &
  \alpha \frac{gv}{2} i \bar b_L \paM t_R \Wm^-
  - \alpha^* \frac{gv}{2} i \paM \bar t_R b_L \Wm^+
  \,, \notag \\
\alpha O_{\bar Du}^{33} + \alpha^* (O_{\bar Du}^{33})^\dagger & \supset &
  \alpha \frac{gv}{2} i \paM  \bar b_L t_R \Wm^-
  - \alpha^* \frac{gv}{2} i  \bar t_R \paM b_L \Wm^+
  \,, \notag \\
\alpha O_{Dd}^{33} + \alpha^* (O_{Dd}^{33})^\dagger & \supset &
  \alpha \frac{gv}{2} i \bar t_L \paM b_R \Wm^+
  - \alpha^* \frac{gv}{2} i \paM \bar b_R t_L \Wm^-
  \,, \notag \\
\alpha O_{\bar Dd}^{33} + \alpha^* (O_{\bar Dd}^{33})^\dagger & \supset &
  \alpha \frac{gv}{2} i \paM  \bar t_L b_R \Wm^+
  - \alpha^* \frac{gv}{2} i  \bar b_R \paM t_L \Wm^-
  \,, \notag \\
\alpha O_{qW}^{33} + \alpha^* (O_{qW}^{33})^\dagger & \supset &
   \sqrt 2 \left[ \RE \alpha \; \paN (\bar b_L \gm t_L)
  + i \,\IM \alpha \; \bar b_L \gm \paND t_L \right] \Wmn^- \notag \\
 & & + \sqrt 2 \left[ \RE \alpha \; \paN (\bar t_L \gm b_L)
  + i \,\IM \alpha \; \bar t_L \gm \paND b_L \right] \Wmn^+
  \,.
\end{eqnarray}
Associated quartic $gWtb$ terms arise only from the redundant operators:
\begin{eqnarray}
\alpha O_{Du}^{33} + \alpha^* (O_{Du}^{33})^\dagger & \supset &
  - \frac{g g_s v}{4} \left[ \alpha\, \bar b_L \la \gmn t_R \, \Wm^-
  +\alpha^* \bar t_R \la \gmn b_L \, \Wm^+ \right] \Gn \,, \notag \\[1mm]
\alpha O_{\bar Du}^{33} + \alpha^* (O_{\bar Du}^{33})^\dagger & \supset &
  \frac{g g_s v}{4} \left[ \alpha \, \bar b_L \la \gmn t_R \, \Wm^-
  +\alpha^* \bar t_R \la \gmn b_L \, \Wm^+ \right] \Gn \,, \notag \\[1mm]
\alpha O_{Dd}^{33} + \alpha^* (O_{Dd}^{33})^\dagger & \supset &
  - \frac{g g_s v}{4}  \left[ \alpha \, \bar t_L \la \gmn b_R \, \Wm^+
  +\alpha^* \bar b_R \la \gmn t_L \,\Wm^- \right] \Gn \,, \notag \\[1mm]
\alpha O_{\bar Dd}^{33} + \alpha^* (O_{\bar Dd}^{33})^\dagger & \supset &
  \frac{g g_s v}{4}  \left[ \alpha \,\bar t_L \la \gmn b_R \,\Wm^+
  +\alpha^* \bar b_R \la \gmn t_L \,\Wm^- \right] \Gn \,, \notag \\[1mm]
\alpha O_{qW}^{33} + \alpha^* (O_{qW}^{33})^\dagger & \supset &
  - \sqrt 2 \, \IM \alpha \, g_s \left[
  \bar b_L \la \gm \gns t_L \Wmn^- + \bar t_L \la \gm \gns b_L \Wmn^+ \right] \Gs \,.
\end{eqnarray}
The contributions from effective operators to the effective $Ztt$ vertex are
\begin{eqnarray}
\alpha O_{\phi q}^{(3,33)} + \alpha^* (O_{\phi q}^{(3,33)})^\dagger & \supset &
 -\RE \alpha \; \frac{gv^2}{2 c_W} \bar t_L \gm t_L \Zm
  \,, \notag \\
\alpha O_{\phi q}^{(1,33)} + \alpha^* (O_{\phi q}^{(1,33)})^\dagger & \supset &
 \RE \alpha \; \frac{gv^2}{2 c_W} \bar t_L \gm t_L \Zm
  \,, \notag \\
\alpha O_{\phi u}^{33} + \alpha^* (O_{\phi u}^{33})^\dagger & \supset &
 \RE \alpha \; \frac{gv^2}{2 c_W} \bar t_R \gm t_R \Zm
  \,, \notag \\
\alpha O_{uW}^{33} + \alpha^* (O_{uW}^{33})^\dagger & \supset &
  \frac{v}{\sqrt 2} c_W  \left[ \RE \alpha \; \bar t \smn t 
  + i\,\IM \alpha \; \bar t \smn \gamma_5 t \right] \Zmn
  \,, \notag \\
\alpha O_{uB\phi}^{33} + \alpha^* (O_{uB\phi}^{33})^\dagger & \supset &
  - \frac{v}{\sqrt 2} s_W  \left[ \RE \alpha \; \bar t \smn t 
  + i\,\IM \alpha \; \bar t \smn \gamma_5 t \right] \Zmn
  \,, \notag \\
\alpha O_{Du}^{33} + \alpha^* (O_{Du}^{33})^\dagger & \supset &
  \frac{gv}{2\sqrt 2 c_W} \left[ \alpha \, i \bar t_L \paM t_R
  - \alpha^* i \paM \bar t_R t_L \right] \Zm
  \,, \notag \\
\alpha O_{\bar Du}^{33} + \alpha^* (O_{\bar Du}^{33})^\dagger & \supset &
  \frac{gv}{2\sqrt 2 c_W} \left[ \alpha \, i \paM \bar t_L t_R
  - \alpha^* i  \bar t_R \paM t_L \right] \Zm
  \,, \notag \\
\alpha O_{qW}^{33} + \alpha^* (O_{qW}^{33})^\dagger & \supset &
   c_W \left[ \RE \alpha \; \paN (\bar t_L \gm t_L)
  + i\,\IM \alpha \; \bar t_L \gm \paND t_L \right] \Zmn
  \,, \notag \\
\alpha O_{qB}^{33} + \alpha^* (O_{qB}^{33})^\dagger & \supset &
   -s_W \left[ \RE \alpha \; \paN (\bar t_L \gm t_L)
  + i\,\IM \alpha \; \bar t_L \gm \paND t_L \right] \Zmn
  \,, \notag \\
\alpha O_{uB}^{33} + \alpha^* (O_{uB}^{33})^\dagger & \supset &
   -s_W \left[ \RE \alpha \; \paN (\bar t_R \gm t_R)
  + i\,\IM \alpha \; \bar t_R \gm \paND t_R \right] \Zmn
  \,.
\end{eqnarray}
Among these operators, the contributions to the $gZtt$ vertex are only from the redundant ones,
\begin{eqnarray}
\alpha O_{Du}^{33} + \alpha^* (O_{Du}^{33})^\dagger & \supset &
- \frac{g g_s v}{4 \sqrt 2 c_W} \left[ \RE \alpha \; \bar t \la \gmn t
+ i \, \IM \alpha \; \bar t \la \gmn \gamma_5 t \right] G_\nu^a \, \Zm \,, \notag \\
\alpha O_{\bar Du}^{33} + \alpha^* (O_{\bar Du}^{33})^\dagger & \supset &
 \frac{g g_s v}{4 \sqrt 2 c_W} \left[ \RE \alpha \; \bar t \la \gmn t
+ i \, \IM \alpha \; \bar t \la \gmn \gamma_5 t \right] G_\nu^a \, \Zm \,, \notag \\
\alpha O_{qW}^{33} + \alpha^* (O_{qW}^{33})^\dagger & \supset &
- \IM \alpha \, g_s c_W \bar t_L \la \gm \gns t_L \, \Gs \Zmn \,, \notag \\[1mm]
\alpha O_{qB}^{33} + \alpha^* (O_{qB}^{33})^\dagger & \supset &
\IM \alpha \, g_s s_W \bar t_L \la \gm \gns t_L \,\Gs \Zmn \,, \notag \\[1mm]
\alpha O_{uB}^{33} + \alpha^* (O_{uB}^{33})^\dagger & \supset &
\IM \alpha \, g_s s_W \bar t_R \la \gm \gns t_R \, \Gs \Zmn \,.
\end{eqnarray}
The operators contributing to the $\gamma tt$ vertex are
\begin{eqnarray}
\alpha O_{uW}^{33} + \alpha^* (O_{uW}^{33})^\dagger & \supset &
  \frac{v}{\sqrt 2} s_W  \left[ \RE \alpha \; \bar t \smn t 
  + i\,\IM \alpha \; \bar t \smn \gamma_5 t \right] \Amn
  \,, \notag \\
\alpha O_{uB\phi}^{33} + \alpha^* (O_{uB\phi}^{33})^\dagger & \supset &
  \frac{v}{\sqrt 2} c_W  \left[ \RE \alpha \; \bar t \smn t 
  + i\,\IM \alpha \; \bar t \smn \gamma_5 t \right] \Amn
  \,, \notag \\
\alpha O_{qW}^{33} + \alpha^* (O_{qW}^{33})^\dagger & \supset &
   s_W \left[ \RE \alpha \; \paN (\bar t_L \gm t_L)
  + i \,\IM \alpha \; \bar t_L \gm \paND t_L \right] \Amn
  \,, \notag \\
\alpha O_{qB}^{33} + \alpha^* (O_{qB}^{33})^\dagger & \supset &
   c_W \left[ \RE \alpha \; \paN (\bar t_L \gm t_L)
  + i\,\IM \alpha \; \bar t_L \gm \paND t_L \right] \Amn
  \,, \notag \\
\alpha O_{uB}^{33} + \alpha^* (O_{uB}^{33})^\dagger & \supset &
   c_W \left[ \RE \alpha \; \paN (\bar t_R \gm t_R)
  + i\,\IM \alpha \; \bar t_R \gm \paND t_R \right] \Amn
  \,.
\end{eqnarray}
The associated $g\gamma tt$ quartic vertices are
\begin{eqnarray}
\alpha O_{qW}^{33} + \alpha^* (O_{qW}^{33})^\dagger & \supset &
- \IM \alpha \, g_s s_W \bar t_L \la \gm \gns t_L \,\Gs \Amn
\,, \notag \\[1mm]
\alpha O_{qB}^{33} + \alpha^* (O_{qB}^{33})^\dagger & \supset &
- \IM \alpha \, g_s c_W \bar t_L \la \gm \gns t_L \,\Gs \Amn
\,, \notag \\[1mm]
\alpha O_{uB}^{33} + \alpha^* (O_{qB}^{33})^\dagger & \supset &
- \IM \alpha \, g_s c_W \bar t_R \la \gm \gns t_R \,\Gs \Amn
\,.
\end{eqnarray}
Finally, the contributions to the $gtt$ interaction are
\begin{eqnarray}
\alpha O_{uG\phi}^{33} + \alpha^* (O_{uG\phi}^{33})^\dagger & \supset &
\frac{v}{\sqrt 2} \left[ \RE \alpha \; \bar t \la \smn t + i \, \IM \alpha \; t \la \smn \gamma_5 t \right] \Gmna
\,, \notag \\
\alpha O_{qG}^{33} + \alpha^* (O_{qG}^{33})^\dagger & \supset &
\left[ \RE \alpha \; \paN (\bar t_L \la \gm t_L) + i \, \IM \alpha \;
\bar t_L \la \gm \paND t_L \right] \Gmna
\,, \notag \\
\alpha O_{uG}^{33} + \alpha^* (O_{uG}^{33})^\dagger & \supset &
\left[ \RE \alpha \; \paN (\bar t_R \la \gm t_R) + i \, \IM \alpha \;
\bar t_R \la \gm \paND t_R \right] \Gmna \,.
\end{eqnarray}
The two redundant operators include several associated quartic vertices with extra $W$, $Z$, $\gamma$ bosons as the ones listed above, as well as new ones with an extra gluon. They are
\begin{eqnarray}
\alpha O_{qG}^{33} + \alpha^* (O_{qG}^{33})^\dagger & \supset &
i \frac{g_s}{2} \left[ \RE \alpha \; \bar t_L [\la,\lb] \gm \gns t_L 
+ i \, \IM \alpha \; t_L \{\la,\lb\} \gm \gns t_L \right] G_\sigma^b \Gmna
\notag \\
& & - \sqrt 2 g \, \IM \alpha \left[ \bar t_L \la \gm \gns b_L W_\sigma^+
+ \bar b_L \la \gm \gns t_L W_\sigma^- \right] \Gmna
\notag \\
& & - \frac{g}{c_W} f_u^L \IM \alpha \;
\bar t_L \la \gm \gns t_L Z_\sigma \Gmna 
\notag \\
& & - 2 Q_t e \, \IM \alpha \;
\bar t_L \la \gm \gns t_L A_\sigma \Gmna 
\,, \notag \\
\alpha O_{uG}^{33} + \alpha^* (O_{uG}^{33})^\dagger & \supset &
i \frac{g_s}{2} \left[ \RE \alpha \; \bar t_R [\la,\lb] \gm \gns t_R 
+ i \, \IM \alpha \; t_R \{\la,\lb\} \gm \gns t_R \right] G_\sigma^b \Gmna
\notag \\
& & - \frac{g}{c_W} f_u^R \, \IM \alpha \;
\bar t_R \la \gm \gns t_R Z_\sigma \Gmna 
\notag \\
& & - 2 Q_t e \, \IM \alpha \;
\bar t_R \la \gm \gns t_R A_\sigma \Gmna 
\,,
\end{eqnarray}
with $f_u^L = 1-4/3 \, s_W^2$, $f_u^R = -4/3 s_W^2$, $Q_t=2/3$.

\section{Operator contributions to top FCN interactions}
\label{sec:b}

In this appendix we give the effective operator contributions to top FCN interactions, also including those operators which are redundant. Contributions for top-charm couplings arise from both combinations $i,j=2,3/3,2$, while for top-up they are obtained setting $i,j=1,3/3,1$. Notice that for $i=j=3$ we can recover the results for $Ztt$ and $\gamma tt$ trivially.
The $Ztu$, $Ztc$ vertices are
\begin{eqnarray}
\alpha O_{\phi q}^{(3,ij)} + \alpha^* (O_{\phi q}^{(3,ij)})^\dagger & \supset &
 - \frac{gv^2}{4 c_W} \left[ \alpha \, \bar u_{Li} \gm u_{Lj}
 + \alpha^* \, \bar u_{Lj} \gm u_{Li} \right] \Zm
  \,, \notag \\
\alpha O_{\phi q}^{(1,ij)} + \alpha^* (O_{\phi q}^{(1,ij)})^\dagger & \supset &
 \frac{gv^2}{2 c_W} \left[ \alpha \, \bar u_{Li} \gm u_{Lj}
 + \alpha^* \, \bar u_{Lj} \gm u_{Li} \right] \Zm
  \,, \notag \\
\alpha O_{\phi u}^{ij} + \alpha^* (O_{\phi u}^{ij})^\dagger & \supset &
 \frac{gv^2}{4 c_W} \left[ \alpha \, \bar u_{Ri} \gm u_{Rj}
 + \alpha^* \, \bar u_{Rj} \gm u_{Ri}  \right] \Zm
  \,, \notag \\
\alpha O_{uW}^{ij} + \alpha^* (O_{uW}^{ij})^\dagger & \supset &
  \frac{v}{\sqrt 2} c_W  \left[ \alpha \, \bar u_{Li}  \smn u_{Rj} 
  + \alpha^* \, \bar u_{Rj} \smn u_{Li} \right] \Zmn
  \,, \notag \\
\alpha O_{uB\phi}^{ij} + \alpha^* (O_{uB\phi}^{ij})^\dagger & \supset &
  - \frac{v}{\sqrt 2} s_W  \left[ \alpha \, \bar u_{Li} \smn u_{Rj} 
  + \alpha^* \, \bar u_{Rj} \smn u_{Li} \right] \Zmn
  \,, \notag \\
\alpha O_{Du}^{ij} + \alpha^* (O_{Du}^{ij})^\dagger & \supset &
  \frac{gv}{2\sqrt 2 c_W} \left[ \alpha \, i \bar u_{Li} \paM u_{Rj}
  - \alpha^* i \paM \bar u_{Rj} u_{Li} \right] \Zm
  \,, \notag \\
\alpha O_{\bar Du}^{ij} + \alpha^* (O_{\bar Du}^{ij})^\dagger & \supset &
  \frac{gv}{2\sqrt 2 c_W} \left[ \alpha \, i \paM \bar u_{Li} u_{Rj}
  - \alpha^* i  \bar u_{Rj} \paM u_{Li} \right] \Zm
  \,, \notag \\
\alpha O_{qW}^{ij} + \alpha^* (O_{qW}^{ij})^\dagger & \supset &
   c_W \left[ \alpha \, \bar u_{Li} \gm \paN u_{Lj}
   + \alpha^* \,  \paN \bar u_{Lj} \gm u_{Li} \right] \Zmn
  \,, \notag \\
\alpha O_{qB}^{ij} + \alpha^* (O_{qB}^{ij})^\dagger & \supset &
   -s_W \left[ \alpha \, \bar u_{Li} \gm \paN u_{Lj}
   + \alpha^* \,  \paN \bar u_{Lj} \gm u_{Li} \right] \Zmn
  \,, \notag \\
\alpha O_{uB}^{ij} + \alpha^* (O_{qB}^{ij})^\dagger & \supset &
   -s_W \left[ \alpha \, \bar u_{Ri} \gm \paN u_{Rj}
   + \alpha^* \,  \paN \bar u_{Rj} \gm u_{Ri} \right] \Zmn
 \,.
\end{eqnarray}
The associated quartic couplings with an extra gluon are
\begin{eqnarray}
\alpha O_{Du}^{ij} + \alpha^* (O_{Du}^{ij})^\dagger & \supset &
- \frac{g g_s v}{4 \sqrt 2 c_W} \left[ \alpha \, \bar u_{Li} \la
  \gmn u_{Rj} +  \alpha^* \, \bar u_{Rj} \la \gmn u_{Li} \right] G_\nu^a \, \Zm \notag \\
\alpha O_{\bar Du}^{ij} + \alpha^* (O_{\bar Du}^{ij})^\dagger & \supset &
 \frac{g g_s v}{4 \sqrt 2 c_W} \left[ \alpha \, \bar u_{Li} \la
  \gmn u_{Rj} +  \alpha^* \, \bar u_{Rj} \la \gmn u_{Li} \right] G_\nu^a \, \Zm \notag \\
\alpha O_{qW}^{ij} + \alpha^* (O_{qW}^{ij})^\dagger & \supset &
i \frac{g_s c_W}{2} \left[ \alpha \bar u_{Li} \la \gm \gns u_{Lj} 
- \alpha^* \bar u_{Lj} \la \gm \gns u_{Li} \right] \Gs \Zmn
 \notag \\
\alpha O_{qB}^{ij} + \alpha^* (O_{qB}^{ij})^\dagger & \supset &
-i \frac{g_s s_W}{2} \left[ \alpha \bar u_{Li} \la \gm \gns u_{Lj} 
- \alpha^* \bar u_{Lj} \la \gm \gns u_{Li} \right] \Gs \Zmn
 \notag \\
\alpha O_{uB}^{ij} + \alpha^* (O_{uB}^{ij})^\dagger & \supset &
-i \frac{g_s s_W}{2} \left[ \alpha \bar u_{Ri} \la \gm \gns u_{Rj} 
- \alpha^* \bar u_{Rj} \la \gm \gns u_{Ri} \right] \Gs \Zmn
 \,.
\end{eqnarray}
For the $\gamma tu$, $\gamma tc$ vertices the corresponding contributions are
\begin{eqnarray}
\alpha O_{uW}^{ij} + \alpha^* (O_{uW}^{ij})^\dagger & \supset &
  \frac{v}{\sqrt 2} s_W  \left[ \alpha \, \bar u_{Li}  \smn u_{Rj} 
  + \alpha^* \, \bar u_{Rj} \smn u_{Li} \right] \Amn
  \,, \notag \\
\alpha O_{uB\phi}^{ij} + \alpha^* (O_{uB\phi}^{ij})^\dagger & \supset &
   \frac{v}{\sqrt 2} c_W  \left[ \alpha \, \bar u_{Li} \smn u_{Rj} 
  + \alpha^* \, \bar u_{Rj} \smn u_{Li} \right] \Amn
  \,, \notag \\
\alpha O_{qW}^{ij} + \alpha^* (O_{qW}^{ij})^\dagger & \supset &
   s_W \left[ \alpha \, \bar u_{Li} \gm \paN u_{Lj}
   + \alpha^* \,  \paN \bar u_{Lj} \gm u_{Li} \right] \Amn
  \,, \notag \\
\alpha O_{qB}^{ij} + \alpha^* (O_{qB}^{ij})^\dagger & \supset &
   c_W \left[ \alpha \, \bar u_{Li} \gm \paN u_{Lj}
   + \alpha^* \,  \paN \bar u_{Lj} \gm u_{Li} \right] \Amn
  \,, \notag \\
\alpha O_{uB}^{ij} + \alpha^* (O_{qB}^{ij})^\dagger & \supset &
   c_W \left[ \alpha \, \bar u_{Ri} \gm \paN u_{Rj}
   + \alpha^* \,  \paN \bar u_{Rj} \gm u_{Ri} \right] \Amn
 \,,
\end{eqnarray}
whereas the associated $g\gamma tu$, $g \gamma tc$ couplings are
\begin{eqnarray}
\alpha O_{qW}^{ij} + \alpha^* (O_{qW}^{ij})^\dagger & \supset &
i \frac{g_s s_W}{2} \left[ \alpha \bar u_{Li} \la \gm u_{Lj} 
- \alpha^* \bar u_{Lj} \la \gm \gns u_{Li} \right] \Gs \Amn
 \notag \\[1mm]
\alpha O_{qB}^{ij} + \alpha^* (O_{qB}^{ij})^\dagger & \supset &
i \frac{g_s c_W}{2} \left[ \alpha \bar u_{Li} \la \gm u_{Lj} 
- \alpha^* \bar u_{Lj} \la \gm \gns u_{Li} \right] \Gs \Amn
 \notag \\[1mm]
\alpha O_{uB}^{ij} + \alpha^* (O_{uB}^{ij})^\dagger & \supset &
i \frac{g_s c_W}{2} \left[ \alpha \bar u_{Ri} \la \gm u_{Rj} 
- \alpha^* \bar u_{Rj} \la \gm \gns u_{Ri} \right] \Gs \Amn
 \,.
\end{eqnarray}
The $gtu$ and $gtc$ vertices can be obtained from the operators
\begin{eqnarray}
\alpha O_{uG\phi}^{ij} + \alpha^* (O_{uG\phi}^{ij})^\dagger & \supset &
\frac{v}{\sqrt 2} \left[ \alpha \bar u_{Li} \la \smn u_{Rj} + \alpha^* \bar u_{Rj} \la \smn u_{Li} \right] \Gmna
\,, \notag \\
\alpha O_{qG}^{ij} + \alpha^* (O_{qG}^{ij})^\dagger & \supset &
\left[ \alpha \bar u_{Li} \la \gm \paN u_{Lj} + \alpha^* \paN \bar u_{Lj} \la \gm u_{Li} \right] \Gmna
\,, \notag \\[1mm]
\alpha O_{uG}^{ij} + \alpha^* (O_{uG}^{ij})^\dagger & \supset &
\left[ \alpha \bar u_{Ri} \la \gm \paN u_{Rj} + \alpha^* \paN \bar u_{Rj} \la \gm u_{Ri} \right] \Gmna
\,.
\end{eqnarray}
The last two also include the associated quartic vertices
\begin{eqnarray}
\alpha O_{qG}^{ij} + \alpha^* (O_{qG}^{ij})^\dagger & \supset &
i \frac{g_s}{2} \left[ \alpha \bar u_{Li} \la \lb  \gm \gns u_{Lj} 
- \alpha^* \bar u_{Lj} \lb \la  \gm \gns u_{Li} \right] G_\sigma^b \Gmna
\notag \\
& & + i \frac{g}{\sqrt 2} \left[ \alpha \bar u_{Li} \la \gm \gns d_{Lj}
- \alpha^* \bar u_{Lj} \la \gm \gns d_{Li} \right]  W_\sigma^+ \Gmna
\notag \\
& & + i \frac{g}{\sqrt 2} \left[ \alpha \bar d_{Li} \la \gm \gns u_{Lj}
- \alpha^* \bar d_{Lj} \la \gm \gns u_{Li} \right]  W_\sigma^- \Gmna
\notag \\
& & + i \frac{g}{2 c_W} f_u^L \left[
\alpha \bar u_{Li} \la \gm \gns u_{Lj} - \alpha^* \bar u_{Lj} \la \gm \gns u_{Li} \right] Z_\sigma \Gmna 
\notag \\[1mm]
& & + i Q_t e \left[ 
\alpha \bar u_{Li} \la \gm \gns u_{Lj} - \alpha^* \bar u_{Lj} \la \gm \gns u_{Li} 
\right] A_\sigma \Gmna
\,, \notag \\[1mm]
\alpha O_{uG}^{ij} + \alpha^* (O_{qG}^{ij})^\dagger & \supset &
i \frac{g_s}{2} \left[ \alpha \bar u_{Ri} \la \lb  \gm \gns u_{Rj} 
- \alpha^* \bar u_{Rj} \lb \la  \gm \gns u_{Ri} \right] G_\sigma^b \Gmna
\notag \\
& & + i \frac{g}{2 c_W} f_u^R \left[
\alpha \bar u_{Ri} \la \gm \gns u_{Rj} - \alpha^* \bar u_{Rj} \la \gm \gns u_{Ri} \right] Z_\sigma \Gmna 
\notag \\[1mm]
& & + i Q_t e \left[ 
\alpha \bar u_{Ri} \la \gm \gns u_{Rj} - \alpha^* \bar u_{Rj} \la \gm \gns u_{Ri} 
\right] A_\sigma \Gmna
\,,
\end{eqnarray}
with $f_u^L = 1-4/3 \, s_W^2$, $f_u^R = -4/3 s_W^2$, $Q_t=2/3$.

\end{document}